%% file: si.tex
\documentclass{article}
\pdfoutput=1
\usepackage {amsmath}
\usepackage {amsfonts}
\usepackage{graphicx}
\usepackage{pdfpages}
\newcommand{\EE}{\ensuremath{\mathbb{E}}}  
\newcommand{\Dir}{\ensuremath{\text{Dir}}} 
\newcommand{\Poi}{\ensuremath{\text{Pois}}} 
\newcommand{\attime}[2]{\ensuremath{#1 ^{(#2)}}}

\newcommand{\del}[1]{\ensuremath{\delta\left(#1\right)}}
\newcommand{\EM}[1]{\ensuremath {\EE \left[ #1 \right]}}
\newcommand{\ef}[1] {\ensuremath{\exp\left( #1\right)}}
\newcommand{\integ}[1]{\ensuremath{ \frac{1}{2N_0}\int_{0}^{#1}\frac{d\tau}{\nu(\tau)}  }}

\usepackage{subfigure}

\title{Supporting Information}
\begin{document}


# The date of interbreeding between Neandertals and modern humans

Sriram Sankararaman[1,2,*], Nick Patterson[2], Heng Li[2], Svante Pääbo[3*] & David Reich[1,2]*

[1]Department of Genetics, Harvard Medical School, Boston, MA, 02115 USA;
[2]Broad Institute of MIT and Harvard, Cambridge, MA, 02142 USA;
[3]Department of Evolutionary Genetics, Max Planck Institute for Evolutionary Anthropology, Leipzig, D-04103 Germany.

* Correspondence to: Sriram Sankaramanan (sankaramaran@genetics.med.harvard.edu), Svante Pääbo (paabo@eva.mpg.de) or David Reich (reich@genetics.med.harvard.edu)

**Abstract**

**Comparisons of DNA sequences between Neandertals and present-day humans have shown that Neandertals share more genetic variants with non-Africans than with Africans. This could be due to interbreeding between Neandertals and modern humans when the two groups met subsequent to the emergence of modern humans outside Africa. However, it could also be due to population structure that antedates the origin of Neandertal ancestors in Africa. We measure the extent of linkage disequilibrium (LD) in the genomes of present-day Europeans and find that the last gene flow from Neandertals (or their relatives) into Europeans likely occurred 37,000-86,000 years before the present (BP), and most likely 47,000-65,000 years ago. This supports the recent interbreeding hypothesis, and suggests that interbreeding may have occurred when modern humans carrying Upper Paleolithic technologies encountered Neandertals as they expanded out of Africa.**

**Author Summary**

**One of the key discoveries from the analysis of the Neandertal genome is that Neandertals share more genetic variants with non-Africans than with Africans. This observation is consistent with two hypotheses: interbreeding between Neandertals and modern humans after modern humans emerged out of African or population structure in the ancestors of Neandertals and modern humans. These hypotheses make different predictions about the date of last gene exchange between the ancestors of Neandertals and modern non-Africans. We estimate this date by measuring the extent of linkage disequilibrium (LD) in the genomes of present-day Europeans and find that the last gene flow from Neandertals into Europeans likely occurred 37,000-86,000 years before the present (BP), and most likely 47,000-65,000 years ago. This supports the recent interbreeding hypothesis, and suggests that interbreeding occurred when modern humans carrying Upper Paleolithic technologies encountered Neandertals as they expanded out of Africa.**

**Introduction**

A much-debated question in human evolution is the relationship between modern humans and Neandertals. Modern humans appear in the African fossil record about 200,000 years ago. Morphological traits typical of Neandertals appear in the European fossil record about 400,000 years ago [1] and disappear about 30,000 year ago. They lived in Europe and western Asia with a range that extended as far east as Siberia [2] and as far south as the middle East. The overlap of Neandertals and modern humans in space and time suggests the possibility of interbreeding. Evidence, both for [3] and against interbreeding [4], have been put forth based on the analysis of modern human DNA. Although mitochondrial DNA from multiple Neandertals has shown that Neandertals fall outside the range of modern human variation [5,6,7,8,9,10], low-levels of gene flow cannot be excluded [10,11,12].

Analysis of the draft sequence of the Neandertal genome revealed that the Neandertal genome shares more alleles with non-African than with sub-Saharan African genomes [13]. One hypothesis that could explain this observation is a history of gene flow from Neandertals into modern humans, presumably when they encountered each other in Europe and the Middle East [13] (Figure 1). An alternative hypothesis is that the findings are explained by ancient population structure in Africa [13,14,15,16], whereby the population ancestral to Neandertal and modern human ancestors was subdivided. If this substructure persisted until modern humans carrying Upper Paleolithic technologies expanded out of Africa so that the modern human population that migrated was genetically closer to Neandertals, people outside Africa today would share more genetic variants with Neandertals that people in sub-Saharan Africa [13,14,15] (Figure 1). Ancient substructure in Africa is a plausible alternative to the hypothesis of recent gene flow. Today, sub-Saharan Africans harbor deep lineages that are consistent with a highly-structured ancestral population [17,18,19,20,21,22,23,24,25,26,27]. Evidence for ancient structure in Africa has also been offered based on the substantial diversity in neurocranial geometry amongst early modern humans [28]. Thus, it is important to test formally whether substructure could explain the genetic evidence for Neandertals being more closely related to non-Africans than to Africans.

A direct way to distinguish the hypothesis of recent gene flow from the hypothesis of ancient substructure is to infer the date for when the ancestors of Neandertals and a modern non-African population last exchanged genes. In the recent gene flow scenario, the date is not expected to be much older than 100,000 years ago, corresponding to the time of the earliest documented modern humans outside of Africa[29]. In the ancient substructure scenario, the date of last common ancestry is expected to be at least 230,000 years ago, since Neandertals must have separated from modern humans by that time based on when the first definitive Neandertals appear in the fossil record of Europe[1].

In present-day human populations, the extent of LD between two single nucleotide polymorphisms (SNPs) shared with Neandertals can be the result of two phenomena. First, there is "non-admixture LD" [30] whose extent reflects stretches of DNA inherited from the ancestral population of Neandertals and modern humans as well as LD that has arisen due to bottlenecks and genetic drift in modern humans since they separated from Neandertals. Second, if gene flow from Neandertals into modern humans occurred, there is "admixture LD"[30], which will reflect stretches of genetic material inherited by modern humans through interbreeding with Neandertals. The extent of LD between single nucleotide polymorphisms (SNPs) shared with Neandertals will thus reflect, at least in part, the time since Neandertals or their ancestors and modern humans or their ancestors last exchanged genes with each other.

The strategy of using LD to estimate dates of gene flow events has been previously been explored by several groups [31,32,33,34,35]. Our methodology is conceptually similar to the methodology developed by Moorjani et al., but is dealing with a more challenging technical problem since the methodology developed by Moorjani et al. is adapted for relatively recent admixtures. In recently admixed populations that have not experienced recent bottlenecks, admixture LD extends over size scales at which non-admixture LD makes a negligible contribution. Thus, one can infer the time of gene flow based on inter-marker spacings that are larger than the scale of non-admixture LD. For older admixtures however (such as may have occurred in the case of Neandertals), non-admixture LD occurs almost at the same size scale as admixture LD. To account for this, we study pairs of markers that are very close to each other, but ascertain them in a way that greatly

minimizes the signals of non-admixture LD while enhancing the signals of admixture LD. Thus, unlike in the case of recent admixtures, non-admixture LD could bias an admixture date obtained using our methods; however, we show using simulations of a very wide set of demographic scenarios that that our marker ascertainment procedure makes the bias so small that our inferences are qualitatively unaffected.

Our methodology is based on the idea that if two alleles, a genetic distance $x$ (expected number of crossover recombination events per meiosis) apart, arose on the Neandertal lineage and introgressed into modern humans at time $t_{GF}$, the probability that these alleles have not been broken up by recombination since gene flow is proportional to $e^{-t_{GF}x}$. The LD across introgressed pairs of alleles is expected to decay exponentially with genetic distance. The rate of decay is informative of the time of gene flow and is robust to demographic events (Appendix A, Supporting Information S1). In practice, we need to ascertain SNPs that, assuming recent gene flow occurred, are likely to have arisen on the Neandertal lineage and introgressed into modern humans. We choose a particular ascertainment scheme and show, using simulations of a number of demographic models, that the exponential decay of LD across pairs of ascertained SNPs provides accurate estimates of the time of gene flow. A second potential source of bias in estimating ancient dates arises from uncertainties in the genetic map. We develop a correction for this bias and show that this correction yields accurate dates in the presence of uncertainties in the genetic map. Combining these various strategies, we are able to obtain accurate estimates of the date of last exchange of genes between Neandertals and modern humans (also see Discussion). This date shows that recent gene flow between Neandertals and modern humans occurred but does not exclude that ancient substructure in Africa also contributes to the LD observed.

**Results**

To study how LD decays with the distance in the genome, we computed the average value, $\overline{D}(x)$, of the measure of linkage disequilibrium $D$ (the excess rate of occurrence of derived alleles at two SNPs compared with the expectation if they were independent[36]) between pairs of SNPs binned by genetic distance $x$ (see Methods). Immediately after the time of last gene flow between Neandertal (or their relatives) and human ancestors, long

range LD is generated, and it is then expected to decay at a constant rate per generation as recombination breaks down the segments shared with Neandertals. Thus, in the absence of new LD-generating events (discussed further below), the $\overline{D}(x)$ statistic across pairs of introgressed alleles is expected to have an exponential decay with genetic distance, and the genetic extent of the decay can thus be interpreted in terms of the time of last shared ancestry between Neandertals (or their relatives) and modern humans (Section S1 and Appendix A in Supporting Information S1).

To amplify the signal of admixture LD relative to non-admixture LD, we restricted our analysis to SNPs where the "derived" allele (the one that has arisen as a new mutation as determined by comparison to chimpanzee) is found in Neandertals and occurs in the tested population at a frequency of <10%. The justification for this frequency threshold is two-fold. First, the signal of Neandertals being more closely related to non-Africans than to Africans is substantially enriched at SNPs below this threshold (Section S1 in Supporting Information S1). Second, under the model of recent gene flow, such SNPs have an increased probability of having arisen due to mutations on the Neandertal lineage; we estimate that about 30% of them will have arisen on the Neandertal lineage under a model of history that we fitted to the data. This ascertainment enriches the class of informative SNPs by a factor of ten (Section S1 in Supporting Information S1). Our simulations show that restricting to this class of SNPs yields accurate estimates of the time of gene flow for a wide range of demographic histories consistent with patterns of human variation (Section S2 in Supporting Information S1).

To assess how useful this statistic is for measuring admixture LD, we performed coalescent simulations of 100 regions of a million base pairs each, for a range of demographic histories chosen to be plausible for Neandertals, West Africans and non-Africans (these histories were constrained by the observed population differentiation between west Africans and Europeans as measured by their $F_{ST}$ and the quantitative extent to which Neandertals share more derived alleles with Europeans than with Africans). The simulation results, which we discuss at length in Section S2 of Supporting Information S1, and summarize in Table 2, show that we obtain accurate and relatively unbiased estimates of the number of generations since admixture (never more than 15%

from the true value) for (1) constant-sized population scenarios, (2) demographic models that include population bottlenecks as well as more recent admixture after the gene flow, (3) hybrid models of ancient structure and recent gene flow, and (4) mutation rates that differ by a factor of 5 from what we use in our main simulations ( see Fig 2). Two other SNP ascertainment schemes yield qualitatively consistent findings but the ascertainment we used provides the most accurate estimates under the range of demographic models considered (Section S5 of Supporting Information S1 and Table 2). The simulations also show that in the absence of gene flow (including in the scenario of ancient subdivision), the dates obtained are always at least 5,000 generations for scenarios of demographic history that match the constraints of real human data. Thus, an empirical estimate of a date much less than 5,000 generations likely reflects real gene flow.

We applied our statistic to data from Pilot 1 of the 1000 Genomes Project, which discovered polymorphisms in 59 West Africans, 60 European Americans, and 60 East Asians (Han Chinese and Japanese from Tokyo) [37]. We binned pairs of SNPs by the genetic distance between them using the deCODE genetic map. We considered all pairs of SNPs that are at most 1cM apart. We computed the average LD over all pairs of SNPs in each bin and fit an exponential curve to the decay of LD (from 0.02-1cM in 0.001cM increments).

Figure 3 shows the extent of LD for pairs of SNPs where both SNPs have a derived allele frequency <10%. This figure shows that the extent of LD is larger in Europeans and East Asians than in West Africans, both when the Neandertal genome carries the derived and when it carries the ancestral allele. Empirical features of these LD decay curves show that, alleles derived in the Neandertal genome, the pattern in Europeans and East Asians is reflecting "admixture LD". LD in West Africans is less extensive when Neandertals carry the derived allele than when they carry the ancestral allele, while the reverse is seen in Eurasians. To understand this, we note that in the absence of gene flow, polymorphic sites where Neandertals carry the derived allele must have arisen from mutations that occurred prior to Neandertal-human divergence so that they are old and recombination will have had a lot of time to break down the LD, while sites where Neandertals carry the ancestral allele mutations will include mutations that have arisen since the Neandertal-

human split and thus LD will be expected to be more extensive, exactly as is seen in West Africans. In contrast, if gene flow occurred, then LD can be greater at sites where Neandertals carry the derived allele as is observed in Europeans and East Asians. This signal persists when we stratify the LD decay curves by the frequency of the ascertained SNPs (Figure S8 in Supporting Information S1). Thus the scale of the LD at these sites must be conveying information about the date of gene flow.

A concern in interpreting the extent of LD in terms of a date is that all available genetic maps (which specify the probability of recombination per generation between all pairs of SNPs) are likely to be inaccurate at the scale of tens of kilobases that is relevant to our analysis. We confirmed that errors in genetic maps can bias LD-based date estimates by simulating a gene flow event 2,000 generations ago using a model in which recombination was localized to hot spots [38] but where the data were analyzed assuming a genetic map that assumed homogeneous recombination rates across the genome. This led to a date of 1,597 generations since admixture. We developed a statistical model of the random errors that relate the true and observed genetic maps (see Methods). The precision of the map is modeled using a scalar parameter $\alpha$. A unit interval of the observed genetic map corresponds to an interval in the true map of expected unit length and variance $1/\alpha$. To validate this error model, we estimated the map error in these simulations ($\alpha$) by comparing the true and the observed genetic maps. Theoretical arguments (Section S3 in Supporting Information S1) show that we can obtain a corrected date ($t_{GF}$) from the uncorrected date in generations ($\lambda$) using the equation $t_{GF} = \alpha(e^{\lambda/\alpha} - 1)$. We applied this correction to obtain a date of 1,926 generations. While this error model appears to provide an adequate description of random errors in a genetic map, it does not account for systematic biases.

To apply this statistical correction to real data, we estimated the error rate $\alpha$ in the genetic map by comparing the genomic distribution of a set of cross-over events from 728 meioses previously detected in a European American Hutterite pedigree [39] to what would be expected if the map were perfect. Unfortunately, the map that we would ideally want to use for estimating the date of Neandertal admixture is not the genetic map that applies to Hutterites today, but the time-averaged genetic map that applied between the

present and the date of gene flow. Obviously, such a map is not available, but we hypothesize that by performing our analyses using a genetic map that is built from samples more closely related to the Hutterite pedigree than the map that we would like to analyze (the deCODE pedigree map built in Icelanders) as well as a genetic map that averages over too long a period of time (the European LD Map, which measures recombination over approximately five hundred thousand years), we can obtain some sense of the robustness of our inferences to uncertainties in how the European genetic map has changed over time.

Table 1 shows the estimates of $\lambda$, $\alpha$ and $t_{GF}$ in Europeans obtained using the two genetic maps. The estimates of $t_{GF}$ are in 1,805-2,043 for both the deCODE and European LD maps. We also estimated $\lambda$ in East Asians using the "East Asian LD map". We find that $\lambda$ in East Asians based on the East Asian LD map is 1,253-1,287, similar to the 1,159-1,183 in Europeans based on the European LD map, although the similarity of the these numbers does not prove the Neandertal genetic material in Europeans and East Asians derives from the same ancestral gene flow event. While a shared ancestral gene flow event is plausible, the gene flow events could in principle have occurred in different places at around the same time [40]. We also cannot reliably estimate the recombination rate correction factor $\alpha$ for the East Asian map because we do not have access to cross-over events in an East Asian pedigree, and hence we do not present an estimate of $t_{GF}$ in East Asians and focus on Europeans in the rest of this paper.

To convert the date estimates in generations to date estimates in years, we use an average generation interval which has been estimated to be 29 in diverse modern hunter gatherer societies as well as in developing and industrialized nation states [41]. We assume a uniform prior probability distribution of generation times between 25 and 33 years per generation for the true value of this quantity and integrate this with the uncertainty of $\lambda$ and $\alpha$, and obtain an estimate of last gene exchange between Neandertals and European ancestors of 47,334-63,146 years for the deCODE map, and 49,021-64,926 years for the European LD Map (95% credible intervals). Taking the conservative union of these ranges, we obtain 47,000-65,000 years BP. In our simulations of ascertainment strategy, we found demographic models that can produce biases in the date estimates that could be

9

as large as 15% (Section S2 in Supporting Information S1). To be conservative, we applied this to the uncorrected dates from each of the maps and then applied the relevant map correction. The union of the resulting intervals leads us to conclude that the true date of gene flow could be as young as 37,000 years BP or as old as 86,000 years BP.

We considered the possibility that our results might be biased by natural selection, which is known to affect patterns of human genetic diversity and to have had a much larger effect closer to genes [42,43]. We estimated the time of gene flow stratifying the SNPs by their distance to the nearest exon, dividing the data into 5 bins such that each bin contained 20% of all the SNPs. Using the deCODE map, we obtain $\lambda$=1,145-1,301 in all bins (Table S8 in Supporting Information S1). This estimate is concordant with the $\lambda$=1,201 obtained without stratification, and suggests that our inferences are not an artifact of LD generated by directional natural selection.

**Discussion**

The date of 37,000-86,000 years BP is too recent to be consistent with the "ancient African population structure" scenario, and strongly supports the hypothesis that at least some of the signal of Neandertals being more closely related to non-Africans than to Africans is due to recent gene flow. These results are concordant with a recent paper by Yang et al [44] that analyzed joint allele frequency spectra, to reject the ancient structure scenario. One possibility that we have not ruled out is that both ancient structure and gene flow occurred in the history of non-Africans. In the simulations reported in Table 2, we show that in this scenario, the ancient structure will tend to make the date estimate older than the truth but by not more than 15%, so that the date of 37,000-86,000 should still provide a valid bound while the less conservative estimate of 47,000-65,000 years should be interpreted as an upper bound on the date of gene flow. Further, we have not been able to differentiate amongst variants of the recent gene flow scenario: a single episode or multiple episodes of gene flow or continuous gene flow over an extended period of time. Our date has a clear interpretation as the time of last gene exchange under a scenario of a single instantaneous gene flow event. In the other scenarios, the date is expected to represent an average over the times of gene flow and should be interpreted as an upper bound on the time of last gene exchange.

While recent gene flow from Neandertals into the ancestors of modern non-Africans is a parsimonious model that is consistent with our results, our analysis cannot reject the possibility that gene flow did not involve Neandertals themselves, but instead populations that were more closely related to Neandertals than any extant populations are today. Thus, the date should be interpreted as the last period of time when genetic material from Neandertals or an archaic population related to Neandertals entered modern humans.

Genetic analyses by themselves offer no indication of where gene flow may have occurred geographically. However, the date in conjunction with the archaeological evidence suggests that the two populations likely met somewhere in Western Eurasia. An attractive hypothesis is the Middle East, where archaeological and fossil evidence indicate that modern humans appeared before 100,000 years ago (as reflected by the modern human remains in Skhul and Qafzeh caves), Neandertals expanded around 70,000 years ago (as reflected for example by the Neandertal remains at Tabun Cave), and modern humans re-appeared around 50,000 years ago [29]. Our genetic date estimates, which have a mostly likely range of 47,000-65,000 years ago (and are confidently below 86,000 years ago), are too recent to be consistent with the appearance of the first fossil evidence of modern humans outside of Africa—that is, our date makes it unlikely that the Neandertal genetic material in modern humans today could arise exclusively due to the gene flow involving the Skhul/Qafzeh modern humans—and instead point to gene flow in a more recent period, possibly when modern humans carrying Upper Paleolithic technologies expanded out of Africa.

**Methods**

**Linkage disequilibrium statistic:** Our procedure computes a statistic based on the LD observed between pairs of SNPs. For all pairs of ascertained SNPs at a genetic distance *x*, we compute the statistic:

$$\overline{D}(x) = \frac{\sum_{(i,j) \in S(x)} D(i,j)}{|S(x)|}$$

Here *S(x)* denotes the set of all pairs of ascertained SNPs that are at a genetic distance *x*, and *D(i,j)* denotes the classic signed measure of linkage disequilibrium, D, at the SNPs *i, j*. The sign of *D(i,j)* is determined by computing D using the derived alleles (defined relative to the chimpanzee base) at SNPs *i* and *j*. Under the gene flow scenario, we expect the contribution of introgression to $\overline{D}(x)$ to have an exponential decay with rate equal to the time of gene flow, provided the gene flow is more recent than the Neandertal-modern human split (Section S1 and Appendix A of Supporting Information S1).

We pick SNPs that contain a derived allele in Neandertal (defined relative to the chimpanzee base) and are polymorphic in the target population with a derived allele frequency <10%. Further details can be found in the Supporting Information, along with simulations exploring the performance of the statistic and demonstrating its properties under various demographic models and ascertainment schemes.

**Preparation of 1000 Genomes Data and alignment to chimpanzee and Neandertal:** We used the 1000 Genomes Pilot 1 genotypes to estimate the LD decay. For each of the panels that were chosen as the target population in our analysis, we restricted our analysis to polymorphic SNPs. The SNPs were polarized relative to the chimpanzee base (*panTro2*).

**Computation of the LD statistic on 1000 Genomes Data:** For the set of ascertained SNPs, we compute $\overline{D}(x)$ as a function of the genetic distance *x* and fit an exponential curve using ordinary least squares for *x* in the range of 0.02cM to 1cM in increments of

0.001 cM. The standard definition of *D* requires the availability of haplotypes. We instead computed *D(i,j)* as the covariance between the genotypes observed at SNPs *i* and *j* [45]. Simulations show that dates estimated using this definition of *D* on unphased genotypes are very similar to the estimates obtained from haplotypes (Section S2.1.1 of Supporting Information S1). We were concerned that the complicated method used in the 1000 Genomes Project for determining genotypes, which involved statistical imputation and probabilistic calling of genotypes based on LD, might in some way be biasing our inferences based on LD. Thus, we also computed *D(i,j)* for all pairs of SNPs that passed our basic filters (SNPs that contain a derived allele in Neandertal and are polymorphic in the target population with derived allele frequency <10% as estimated from the reads) by computing LD directly from the reads, again using the SAMtools package[46], and obtain qualitatively consistent results (Section S7 of Supporting Information S1). Further, simulations to mimic the low power to call rare SNPs in the 1000 genomes data show that our estimates are not sensitive to the deficit of rare alleles (Section S6 of Supporting Information S1).

**Correction for error in the genetic map:** We have a genetic map G defined on *m* markers. Each of the *m-1* intervals is assigned a genetic distance $g_i$, *i=1,..m-1*. These genetic distances provide a prior distribution for the true underlying (unobserved) genetic distances $Z_i$. A reasonable prior on each $Z_i$ is then:

$$Z_i \sim \Gamma(\alpha g_i, \alpha)$$

where $\alpha$ is a parameter that is specific to the map. This implies that the true genetic distance $Z_i$ has mean $g_i$ and variance $g_i/\alpha$. Thus, large values of $\alpha$ correspond to a more precise map. A motivation for the choice of the gamma prior over $Z_i$ is that this prior has the key invariance property $Z_1+Z_2 \sim \Gamma(\alpha(g_1+g_2), \alpha)$. Thus, $\alpha$ is a property of the map and not of the specific markers used.

Given this prior on the true genetic distances, fitting an exponential function to pairs of markers at a given observed genetic distance *g* involves integrating over the exponential function evaluated at the true genetic distances given observed genetic distance *g*, that is:

13

$$\mathbb{E}\left[\exp\left(-t_{GF}Z\right)|g\right] = \exp\left(-\lambda g\right)$$

where $\lambda$ is the rate of decay of $\overline{D}(g)$ as a function of the observed genetic distance $g$ and can be estimated from the data as described in the previous section, $t_{GF}$ denotes the true time of the gene flow and the expectation is over the unobserved true genetic distance $Z$. We can use this equation to solve for $t_{GF}$ as (see Appendix B, SI):

$$t_{GF} = \alpha \left(\exp\left(\frac{\lambda}{\alpha}\right) - 1\right)$$

To estimate $\alpha$ for a given genetic map, we propose a statistical model that relates the true unobserved genetic map to the observed map and to crossover events found in a pedigree. We estimate the posterior distribution of $\alpha$ by Gibbs sampling (Section S3 of Supporting Information S1).

**Uncertainty in the date estimate taking into account all sources of error:** To obtain estimates of the time of gene flow taking into account all sources of error, we formulated a Bayesian model that relates $\lambda$, $t_{GF}$, and $y_{GF}$ (the time in years) (Section S4 of Supporting Information S1) to the observed LD decay curve.

Further, we assume a uniform prior distribution on the number of years per generation of 25-33 years, based on a recent survey of generation intervals, which are similar in diverse hunter-gatherer societies and in undeveloped as well as industrialized nation states.

Assuming a flat prior on each of $\lambda$, $t_{GF}$, and $y_{GF}$, we use Gibbs sampling to obtain samples from the posterior distributions of each of these parameters. We then report the posterior mean and 95% Bayesian credible intervals.

**Availability:** We will make the data and programs available at
[http://genetics.med.harvard.edu/reichlab/Reich_Lab/Datasets.html](http://genetics.med.harvard.edu/reichlab/Reich_Lab/Datasets.html) on publication.

# Acknowledgments

# Table 1

| Map | $\lambda$ (95% credible interval) | $t_{GF}$ (generations) (95% credible interval) | $y_{GF}$ (years) (95% credible interval) |
|---|---|---|---|
| Decode | 1,179-1,233 | 1,805-1,993 | 47,334-63,146 |
| European LD | 1,159-1,183 | 1,881-2,043 | 49,021-64,926 |

Note: The table gives the admixture dates for Europeans. For East Asians we obtain $\lambda$=1,253-1,287, although no valid conversion to $t_{GF}$ is possible without an East Asian pedigree map and hence we focus on the results for Europeans in this study.

# Table 2

| Demography | Fst (Y,E) | D(Y,E,N) | Ascertainment 0 | Ascertainment 1 | Ascertainment 2 |
|---|---|---|---|---|---|
| **No ancient structure and no gene flow** | | | | | |
| NGF I | 0.15 | 0 | 8847±126 | 7940±257 | 10206±280 |
| NGF II | 0.15 | 0 | 5800±164 | 7204±356 | 11702±451 |
| **Ancient structure** | | | | | |
| AS I | 0.15 | 0.045 | 10128±127 | 8162±107 | 8861±110 |
| AS II | 0.19 | 0.046 | 5070±397 | 6349±327 | 7570±433 |
| **Gene flow 2,000 generations ago** | | | | | |
| RGF II | 0.15 | 0.041 | 1987±48 | 1693±39 | 1960±43 |
| RGF III | 0.14 | 0.043 | 1776±87 | 1643±98 | 2272±102 |
| RGF IV | 0.15 | 0.04 | 2023±56 | 1751±36 | 1995±38 |
| RGF V | 0.07 | 0.04 | 2157±22 | 2094±22 | 2105±22 |
| RGF VI | 0.15 | 0.04 | 2102±36 | 1814±35 | 2029±38 |
| **Hybrid models of ancient structure and gene flow 2,000 generations ago** | | | | | |
| HM I | 0.18 | 0.03 | 2174±40 | 2057±30 | 2228±38 |
| HM II | 0.12 | 0.04 | 2226±39 | 2049±30 | 2100±30 |
| HM III | 0.13 | 0.04 | 2137±34 | 2040±29 | 2124±30 |
| HM IV | 0.18 | 0.06 | 2153±36 | 2038±34 | 2187±35 |
| **Gene flow 2,000 generations ago along with a varying mutation rate** | | | | | |
| $\mu = 1\times10^{-8}$/bp/gen. | 0.11 | 0.04 | 2141±41 | 1847±35 | 1969±36 |
| $\mu = 5\times10^{-8}$/bp/gen. | 0.11 | 0.04 | 2134±41 | 1833±29 | 1951±29 |

The table presents estimates of the time of gene flow for different demographic models and mutation rates as well as different ascertainments. The main classes of models are a) NGF: No gene flow in a randomly mating population; b) AS: Ancient structure, c) RGF : Recent (2,000 generation ago) gene flow from Neandertals (N) into European ancestors (E), d) HM: Hybrid models with ancient structure and recent gene flow and e) Mutation rates that are set to $1\times10^{-8}$/bp/generation and $5\times10^{-8}$/bp/generation. The parameters of the models were chosen to match observed $F_{ST}$ between Africans (Y) and Europeans (E) and to match the observed D-statistics of Africans and Europeans relative to Neandertal D(Y,E;N). In all models that involve recent gene flow, the time of gene flow was set to 2,000 generations. Our estimator of the time of gene flow provides accurate estimates of the time of gene flow for a wide range of demographic and mutational parameters. More details on the models and the ascertainments are in Fig 2, SI S2 and S5.

# Figure Legends

**Figure 1: Linkage disequilibrium patterns expected due to recent gene flow and ancient structure.** (A) In the case of recent gene flow from Neandertals (NEA) into the ancestors of non-Africans (CEU) but not into the ancestors of Africans (YRI), we expect long range LD at sites where Neandertal has the derived allele, and this expectation of admixture generated LD is verified by computer simulation as shown in the right of the panel along with a fitted exponential decay curve. (B) In the case of ancient structure, we expect short range LD, reflecting the >230,000 years since Neandertals and non-Africans derived from a shared ancestral population, and this expectation is also verified by simulation.

**Figure 2: Classes of demographic models relating Africans (Y), Europeans (E) and Neandertals (N)**. a) Recent gene flow but no ancient structure. RGF I has no bottleneck in E. RGF II has a bottleneck after E while RGF VI has a bottleneck after E. RGF IV and V have constant population sizes of $N_e=5000$ and $N_e=50000$ respectively. b) Ancient structure but no recent gene flow. AS I has a constant population size while AS II has a recent bottleneck in E. c) Neither ancient structure nor recent gene flow. NGF I has a constant population size while NGF II has a recent bottleneck in E. d),e) Ancient structure + Recent gene flow. HM IV consists of continuous migration in the Y-E ancestor and the Y-E-N ancestor while HM I consists of continuous migration only in the Y-E ancestor. HM II consist of a single admixture event in the ancestor of E while HM III also models a small population size in one of the admixing populations.

**Figure 3: Decay of LD for SNPs with minor allele frequency <10%**. (A, B) Real data for European Americans and East Asians shows longer range LD when the Neandertal genome has the derived allele (left) than when it has the ancestral allele (right). This is as expected due to gene flow from Neandertal, but is not expected in the absence of gene flow. In other words, the fact that LD conditional on Neandertals having the derived allele is longer than LD when Neandertal does not is proof that the pattern we are observing among ascertained SNPs is reflecting the complex historical relationship between non-African modern humans and Neandertals, the signal we care about here, and not demographic events that solely involve the ancestors of non-Africans. The scale of the LD decay ($1/e$ drop of the fitted exponential curve) is shown in the top right of each panel based on the deCODE genetic distance. (In Figure S8 of supporting Information S1, we show that this signal persists when stratified into narrow allele frequency bins.) (C) In West Africans the pattern is qualitatively different such that when Neandertal is derived at both SNPs, LD decays more quickly than when Neandertal is ancestral at both SNPs, as expected in the absence of gene flow (without gene flow, the derived allele is always expected to be older so LD is expected to have had more time to break down).


\maketitle
\tableofcontents
\input{statistic.tex}
\clearpage
\input{sim-results.tex}

\clearpage
\input{map-error.tex}
\clearpage
\input{uncertainty.tex}
\clearpage
\input{ascertain.tex}
\clearpage
\input{more-results.tex}
\clearpage
\appendix
\input{theory.tex}
\clearpage
\input{appb.tex}
\clearpage
\bibliographystyle{unsrt}
\bibliography{paper}
\end{document}

%% file: statistic.tex
\section{Statistic for dating}
\label{sec:statistic}
A number of methods have been proposed to infer the demographic history and thus the population divergence times of closely-related species using multi-locus genotype data (see~\cite{pinho} and references therein). In this work, we seek to directly estimate the quantity of interest, \emph{i.e}, the time of gene flow, by devising a statistic that is robust to demographic history. Our statistic is based on the pattern of LD decay due to admixture that we observe in a target population. The use of LD decay to test for gene flow is not entirely new (~\cite{machado, moorjani}). ~\cite{machado} devised an LD-based statistic to test the hypotheses of recent gene flow vs ancient shared variation. ~\cite{moorjani} devised a statistic that used the decay of LD to obtain dates of recent gene flow events. The main challenge in our work is the need to estimate extremely old gene flow dates (at least $10000$ years BP) while dealing with the uncertainty in recombination rates.

\subsection{Statistic}
Consider three populations $YRI,CEU$ and $Neandertal$, which we denote $(Y,E,N)$. We want to estimate the date of last exchange of genes between $N$ and $E$. In our demographic model, ancestors of $(Y,E)$ and $N$ split $t_{NH}$ generations ago and $Y$ and $E$ split $t_{YE}$ generations ago. Assume that the gene flow event happened $t_{GF}$ generations ago with a fraction $f$ of individuals from $N$. We have SNP data from several individuals in $E$ and $Y$ as well as low-coverage sequence data for $N$.

\begin{enumerate}
\item Pick SNPs according to an ascertainment scheme discussed below. 
\item For all pairs of sites $S(x)=\{(i,j)\}$ at genetic distance $x$, consider the statistic $\overline{D}(x) = \frac{\sum_{(i,j)\in S(x)}  D(i,j)}{|S(x)|}$. Here $D(i,j)$ is the classic signed measure of LD that measures the excess rate of occurence of derived alleles at two SNPs compared to the expectation if they were independent~\cite{lewontin}.  
\item If there was admixture and if our ascertainment picks pairs of SNPs that arose in Neandertal and introgressed (\emph{i.e.}, these SNPs were absent in $E$ before gene flow), we expect $\overline{D}(x)$ to have an exponential decay with rate given by the time of the admixture because $\overline{D}(x)$ is a consistent estimator of the expected value of $D$ at genetic distance $x$. We can show that, under a model where gene flow occurs at a time $t_{GF}$ and the truly introgressed alleles evolve according to Wright-Fisher diffusion, this expected value has an exponential decay with rate given by $t_{GF}$. Importantly, changes in population size do not affect the rate of decay although imperfections of the ascertainment scheme will affect this rate (see Appendix ~\ref{sec:theory}  for details).
\end{enumerate}

\begin{figure}
\begin{center}
\includegraphics[width=.6\textwidth]{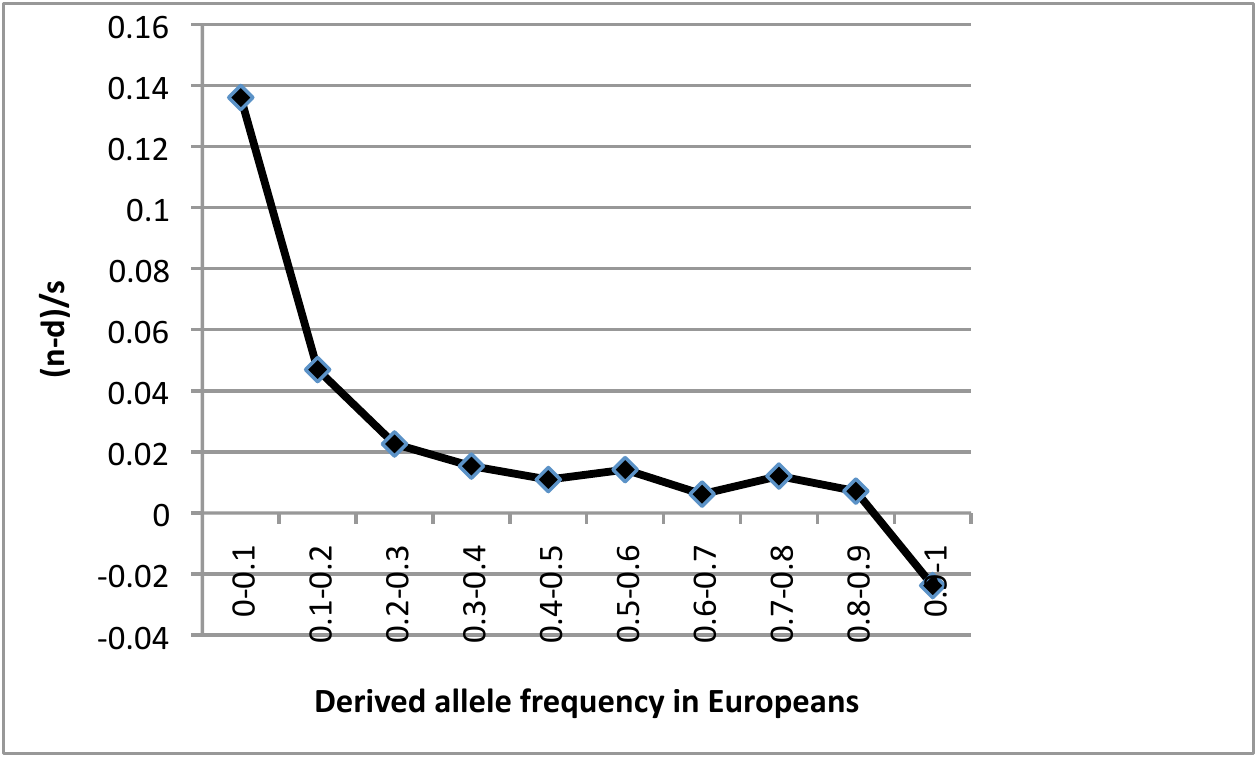}
\end{center}
\caption{The fraction of SNPs $s$ where there is an excess of Neandertal derived alleles $n$ over Denisova derived alleles $d$ as a function of the derived allele frequency in Europeans.}
\label{fig:excess}
\end{figure}

%
%
We pick SNPs that are derived in $N$ (at least one of the reads that maps to the SNP carries the derived allele), are polymorphic in $E$ and have a derived allele frequency in $E < 0.1$. This ascertainment enriches for SNPs that arose in the $N$ lineage and introgressed into $E$ (in addition to SNPs that are polymorphic in the $NH$ ancestor and are segregating in the present-day population). We chose a cutoff of $0.10$ based on an analysis that computes the excess of the number of sites where Neandertal carries the derived allele compared to the number of sites where Denisova carries the derived allele stratified by the derived allele frequency in European populations ($\frac{(n-d)}{s}$ where $s$ is the total number of polymorphic SNPs in Europeans). Given that Denisova and Neandertal are sister groups, we expect these numbers to be equal in the absence of gene flow. The magnitude of this excess is an estimate of the fraction of Neandertal introgressed alleles. Below a derived allele frequency cutoff of $0.10$ in Europeans, we see a significant enrichment of this statistic indicating that it is this part of the spectrum that is most informative for this analysis (see Figure~\ref{fig:excess}). 

To further explore the properties of this ascertainment scheme, we performed coalescent simulations under the RGF II model discussed in Section~\ref{sec:sims}. We computed the fraction of ascertained SNPs for which the lineages leading to the derived alleles in $E$ coalesce with the lineage in $N$ before the split time of Neandertals and modern humans. This estimate provides us a lower bound on the number of SNPs that arose as mutations on the $N$ lineage. We estimate that $30\%$ of the ascertained SNPs arose as mutations in $N$ leading to about 10-fold enrichment over the background rate of introgressd SNPs which has been estimated at $1-4\%$~\cite{green.science.2009}.

We also explored other ascertainment schemes in Section~\ref{sec:ascertain}. 

For the set of ascertained SNPs, we compute $\overline{D}(x)$ as a function of the genetic distance $x$  and fit an exponential curve using ordinary least squares for $x$ in the range of $0.02$ cM to $1$ cM in increments of $10^{-3}$ cM. The standard definition of $D$ requires haplotype frequencies. To compute $D_{i,j}$ directly from genotype data, we estimated $D_{i,j}$ as the covariance between the genotypes observed at SNPs $i$ and $j$~\cite{weir}. We tested the validity of using genotype data on our simulations in Section~\ref{sec:sims}.

\subsection{Preparation of 1000 genomes data}
We used the individual genotypes that were called as part of the pilot 1 of the 1000 genomes project~\cite{1kg} to estimate the LD decay. For each of the panels that were chosen as the target population in our analysis, we restricted ourselves to polymorphic SNPs. The SNPs were polarized relative to the chimpanzee base(PanTro2). 

%% file: sim-results.tex
\section{Simulation Results}
\label{sec:sims}
\begin{figure}
\centering
\subfigure[RM: Recent gene flow]{
\includegraphics[width=.3\textwidth]{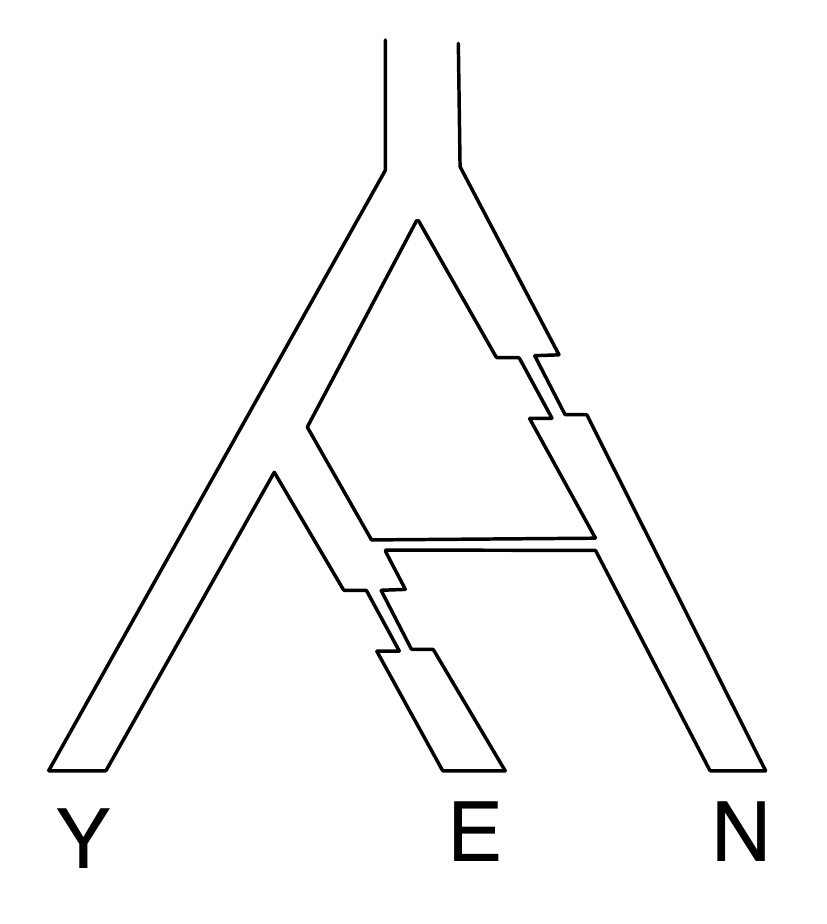}
}
\subfigure[AS: Ancient structure]{
\includegraphics[width=.3\textwidth]{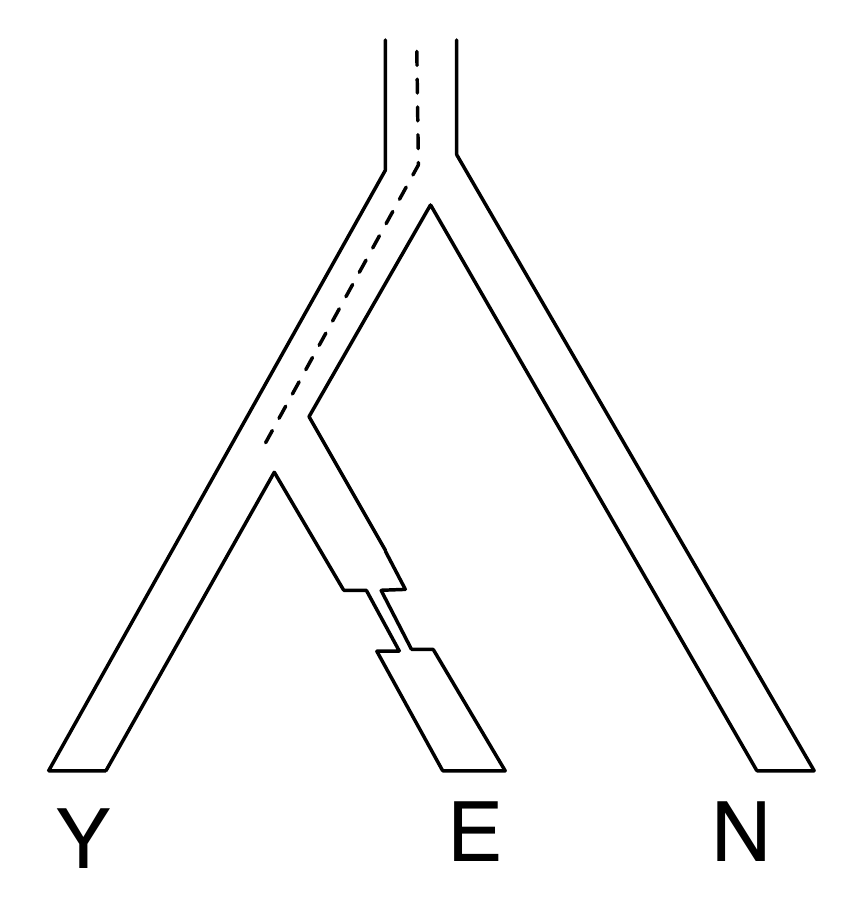}
}

\subfigure[NGF: No gene flow]{
\includegraphics[width=.3\textwidth]{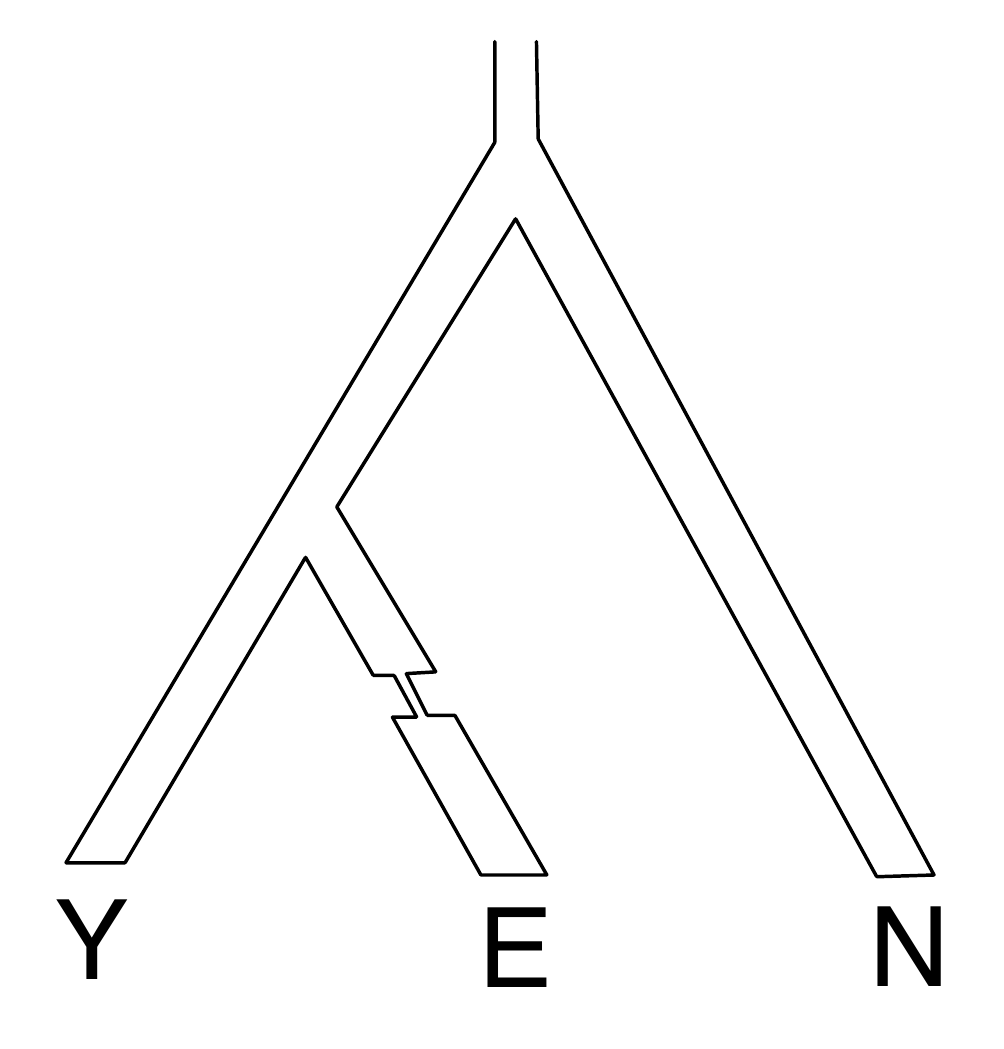}
}
\subfigure[HM: Hybrid model]{
\includegraphics[width=.3\textwidth]{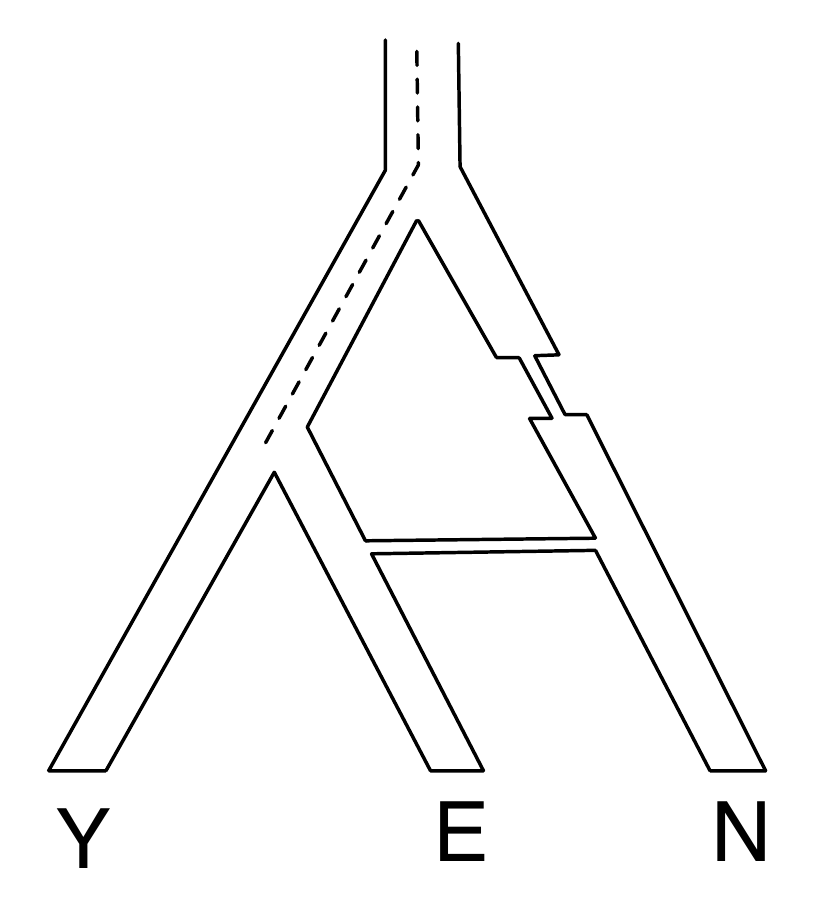}
}
\subfigure[HM: Hybrid model]{
\includegraphics[width=.3\textwidth]{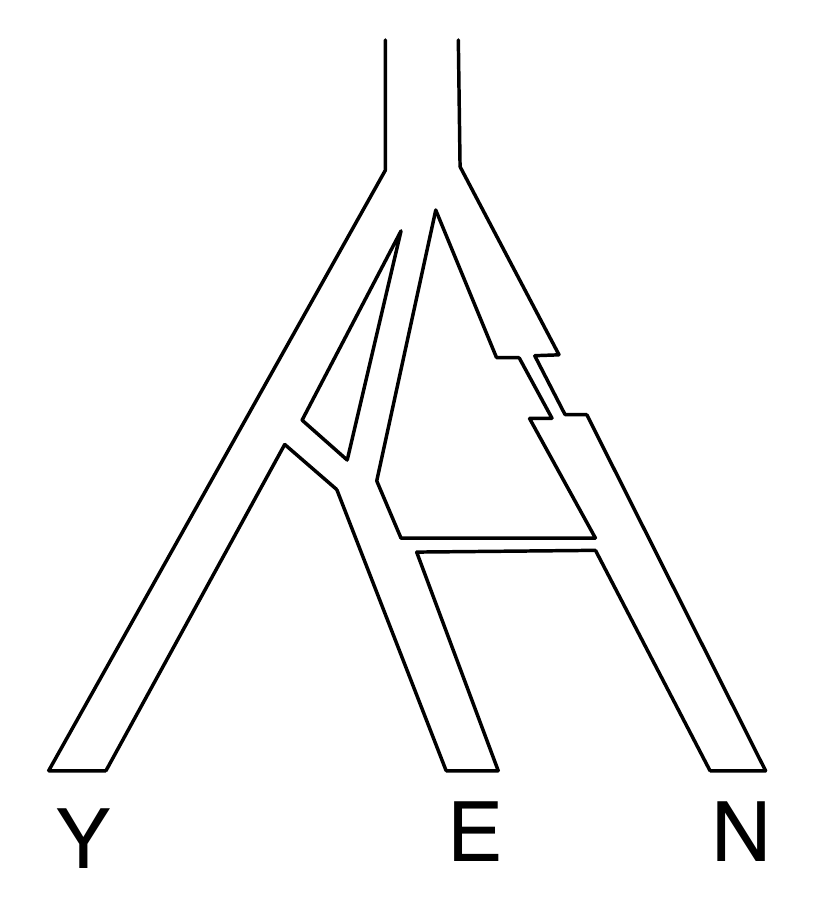}
}

\label{fig:demo}
\caption{Classes of demographic models : a) Recent gene flow but no ancient structure. RGF I has no bottleneck in $E$. RGF II has a bottleneck after $E$ while RGF VI has a bottleneck after $E$. RGF IV and V have constant population sizes of $N_e=5000$ and $N_e=50000$ respectively. b) Ancient structure but no recent gene flow. AS I has a constant population size while AS II has a recent bottleneck in $E$. c) Neither ancient structure nor recent gene flow. NGF I has a constant population size while NGF II has a recent bottleneck in $E$. d),e) Ancient structure + Recent gene flow. HM IV consists of  continuous migration in the $Y-E$ ancestor and the $Y-E-N$ ancestor while HM I consists of continuous migration only in the $Y-E$ ancestor. HM II consist of a single admixture event in the ancestor of $E$ while HM III also models a small population size in one of the admixing populations.}
\end{figure}

To test the robustness of our statistic, we performed coalescent-based simulations under demographic models that included recent gene flow, ancient structure and neither gene flow nor ancient structure. The classes of demographic models are shown in Figure~\ref{fig:demo}

\subsection{Recent gene flow}
\subsubsection {RGF I}
In our first set of simulations, we generated $100$ independent $1$ Mb regions under a simple demographic model of gene flow from Neandertals into non-Africans. We set $t_{NH}=10000$, $t_{YE}=5000$. All effective population sizes are $10000$. The fraction of gene flow was set to $0.03$. We simulated $100$ $Y$ and $E$ haplotypes respectively and $1$ $N$ haplotype. While we simulate a single haploid Neandertal, the sequenced Neandertal genome consists of DNA from 3 individuals. Hence, the reads obtained belong to one of $6$ chromosomes. However, our statistic relies on the Neandertal genome sequence only to determine positions that carry a derived allele. We do not explicitly leverage any pattern of LD from this data. In our simulations, two SNPs at which Neandertal carries the derived allele necessarily lie on a single chromosome and ,hence, are more likely to be in LD than two similar SNPs in the sequenced Neandertals. However, the genetic divergence across the sequenced Vindija bones is quite low (~\cite{reich.science.2010} estimates the average genetic divergence to be about $6000$ years) and so, we do not expect that this makes a big difference in practice.

We simulated $100$ random datasets varying $t_{GF}$ from $0$ to $4500$. Figure ~\ref{fig:date-simple} shows the estimated $t_{GF}$ tracks the true $t_{GF}$ across the range of values of $t_{GF}$. As $t_{GF}$ increases, the variance of our estimates increases -- a result of the increasing influence of the non-admixture LD on the signals of ancient admixture LD. These results are encouraging given that our estimates were obtained using only about $\frac{1}{30}^{th}$ of the data that is available in practice. Further, to test the validity of the use of genotype data, we also computed Pearson's correlation $r$ of estimates of $t_{GF}$ obtained from genotype data to estimates obtained from haplotype data and we estimated these correlations to range from $0.89$ to $0.96$ across different true $t_{GF}$ (see Table~\ref{table:cor}).

\subsubsection {RGF II}
We assessed the effect of demographic changes since the gene flow on the estimates of the time of gene flow. We used $t_{NH}=10000$, $t_{YE}=2500$ and $t_{GF}=2000$. The fraction of gene flow was set to $0.03$. We simulated a bottleneck at $1020$ generations of duration $20$ generations in which the effective population size decreased to $100$. We also simulated a $120$ generation bottleneck in Neandertals from  $3120$ generations in which the effective population size decreased to $100$. These parameters were chosen so that $F_{st}$ between $Y$ and $E$ and the D-statistic $D(Y,E,N)$ match the observed values~\cite{green.science.2009} (the value of the D-statistic $D(Y,E,N)$ depends on the probability of a European lineage entering the Neandertal population and coalescing with a Neandertal lineage before $t_{NH}$ and could have been fit to the data by also adjusting $f$ or $t_{NH}$) . We see in Table~\ref{table:estimates} that the estimated time remains unbiased.  

\subsubsection{ RGF III}
We used a version of the demography used in ~\cite{wall.mbe.2009} modified to match the $F_{st}$ between $Y$ and $E$ and the D-statistics $D(Y,E,N)$. In this setup, $t_{NH}=14400$, $t_{YE}=2400$ ,$t_{GF}=2000$, $f=0.03$. Effective population sizes are $10000$ in the $E$, $YE$ ancestor, $NH$ ancestor, and $10^6$ in modern day $Y$. Modern day $Y$ underwent exponential growth from a size of $10000$ over the last $1000$ generations. $Y$ and $E$ exchange genes after the split at a rate of $150$ per generation. $E$ underwent a bottleneck starting at $1440$ generations that lasted $40$ generations and had an effective population size of $320$ during the bottleneck. We again generated $100$ independent $1$ Mb regions under this demography.

Table~\ref{table:estimates} shows that the estimates now have a small downward bias.

\subsubsection{ RGF IV,V, VI}
This is the same as RGF II but instead of a bottleneck we simulated a constant $N_e$ in population $E$ since gene flow. $N_e$ was set to $5000$ (RGF IV) and  $50000$ (RGF V). RGF VI places the bottleneck before the gene flow ( the bottlenck begins at $2220$ generations, has a duration of $20$ generations in which the effective population size decreased to $100$). Table~\ref{table:estimates} shows that the estimates remain accurate in these settings.

\subsection{Ancient structure}
We examined if ancient structure could produce the signals that we see. We considered a demography (AS I) in which an ancestral panmictic population split to form the ancestors of modern-day $Y$ and another ancestral population $15000$ generations ago. The two populations had low-level gene flow (with population-scaled migration rate of $5$ into $Y$ and $2$ leaving $Y$). The latter population split $9000$ generations ago to form $E$ and $N$. $E$ and $Y$ continued to exchange genes at a low-level down to the present (at a rate of $10$). These parameters were again chosen to match the observed $F_{st}$ between $Y$ and $E$ and $D(Y,E,N)$. Given the longer time scales (here and in the no gene flow model discussed next), we fit an exponential to our statistic over all distances up to $1$ cM. We see from Table~\ref{table:estimates} that we estimate average times of around $10000$ generations.

We also modified the above demography so that $E$ experienced a $20$ generation bottleneck that reduced their $N_e$ to $100$ that ended $1000$ generations ago (AS II). Table ~\ref{table:estimates} shows that our estimates are biased downwards significantly to around $5000$ generations. Nevertheless, we also observe that the magnitude of the exponential, \emph{i.e.}, its intercept, is also decreased. We also considered increasing the duration of the bottleneck but observed that the magnitude of the exponential decay is further diminished and becomes exceedingly noisy.

\subsection {No gene flow}
We also considered a simple model of population splits without any gene flow from $N$ to $E$ (NGF I). We used $t_{NH}=10000$, $t_{YE}=2500$. To investigate if the observed decay of LD could be a result of variation in the effective population size, we also considered a variation (NGF II) with a bottleneck in $E$ at $1020$ generations of duration $20$ generations in which the effective population size decreased to $100$. Table~\ref{table:estimates} shows that our statistic estimates a date of around $8800$ generations in NGF I which is reduced to around $5800$ due to the bottleneck.

Our simulation results show that the LD-based statistic can accurately detect the timing of recent gene flow under a range of demographic models. On the other hand, population size changes in the target population can result in relatively recent dates when there is no gene flow or in the context of ancient structure. This motivated us to explore alternate ascertainment strategies in Section~\ref{sec:ascertain}.

\subsection {Hybrid Models}
These models consist of a recent gene flow from $N$ to $E$ but also simulate structure in the ancestral population of $E$ \emph{i.e.}, in $E$ before gene flow. We would like to explore how ancestral structure affects estimates of the time of last gene exchange. In all these models, we set $t_{GF}=2000, f=0.03$. We consider several such models:
\begin{enumerate}
\item HM I: This is RGF II with no bottleneck in $E$. Instead, the ancestral population of $E$ and $Y$ is structured with the ancestors of $E$ and $Y$ exchanging migrants at a population-scaled rate of $5$. This structure persists from $t_{NH}=10000$ to $t_{YE}=2500$ generations. The population ancestral to modern humans and Neandertal is panmictic.
\item HM II: Similar to HM I. The ancestral population of $E$ is a $0.8:0.2$ admixture of two populations, $E_1$ and $E_2$, just prior to $t_{GF}$. $E_1$ split from $Y$ at time $t_{YE}$ while $E_2$ split from $Y$ at time $t_{NH}$ (resulting in a trifurcation at $t_{NH}$). .
\item HM III: Like in HM II, the ancestral population of $E$ is admixed. $E_2$, in this model, has $N_e=100$ throughout its history.
\item HM IV: This is similar to HM I. The structure in the ancestor of $E$ and $Y$ persists in the Neandertal-modern human ancestor. The ancestor now consists of two subpopulations exchanging migrants at a population-scaled rate of $5$ till $15000$ generations when the population becomes panmictic. $N$ diverges from the subpopulation that is ancestral to $E$ at time $t_{NH}$.
\end{enumerate}
Table~\ref{table:estimates} shows that $t_{GF}$ is accurately estimated, albeit with a small upward bias, under these hybrid demographic models.

\subsection{ Effect of the mutation rate}
Mutation rate has an indirect effect on our estimates -- the mutation rate affects the proportion of ascertained SNPs that are likely to be introgressed. We varied the mutation rate to $1\times10^{-8}$ and $5\times10^{-8}$ in the RGF II model with no European bottleneck and again obtained consistent estimates (Table~\ref{table:estimates}).

\begin{figure}
\begin{center}
\includegraphics[height=.4\textheight]{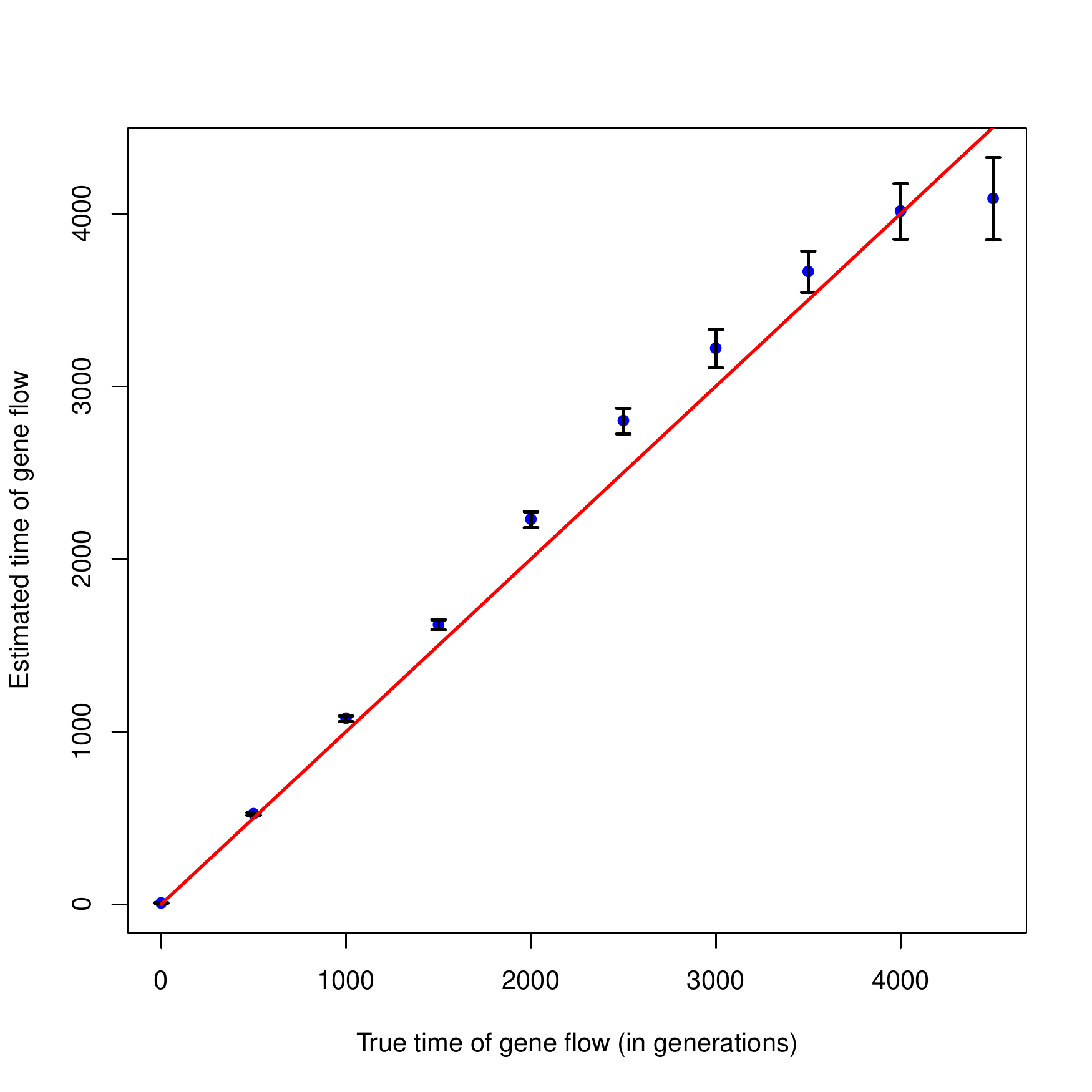}
\end{center}
\caption{Estimates of $t_{GF}$ as a function of true $t_{GF}$ for RGF I: We plot the mean and $2\times$ standard error of the estimates of $t_{GF}$ from $100$ independent simulated datasets using ascertainment 0. The estimates track the true $t_{GF}$ though the variance increases for more ancient gene flow events.}
\label{fig:date-simple-0-1}
\end{figure}

%
%
%

\begin{table}
\begin{center}
\begin{tabular}{lllll}
\hline
Demography&$F_{st}(Y,E)$&$D(Y,E,N)$&\\
\hline
RGF II &0.15 & 0.041 &1987$\pm$48\\
RGF III&0.14 & 0.043 &1776$\pm$87\\
RGF IV &0.15 &0.04& 2023 $\pm$ 56\\
RGF V & 0.07 &0.04&2157$\pm$22\\
RGF VI & 0.15 & 0.04 & 2102 $\pm$ 36\\
AS I & 0.15 & 0.045 & 10128$\pm$127\\
AS II &0.19&0.046&5070$\pm$397\\
NGF I &0.15 &-21$\times10^{-5}$&8847$\pm$	126\\
NGF II& 0.15 & 9$\times10^{-5}$ &5800$\pm$   164\\
HM I &0.18&0.03&2174$\pm$40\\
HM II &0.12&0.04&2226$\pm$39\\
HM III &0.13&0.04&2137$\pm$34\\
HM IV &0.18&0.06&2153$\pm$36\\
\hline
Mutation rate&$F_{st}(Y,E)$&$D(Y,E,N)$& \\
$1\time 10^{-8}$&0.11&0.04&2141$\pm$41\\
$5\times10^{-8}$&0.11&0.04&2134$\pm$41\\
\end{tabular}
\end{center}
\caption{Estimates of the time of gene flow for different demographies and mutation rates.}
\label{table:estimates}
\end{table}

\begin{table}
\begin{center}
\begin{tabular}{ll}
\hline
True $t_{GF}$&Pearson's correlation \\
\hline
0&	0.960918 \\
500&	0.9421455 \\
1000&	0.9335201 \\
1500&	0.9429699 \\
2000&	0.9339092 \\
2500&	0.9464859 \\
3000&	0.9378165 \\
3500&	0.8903148 \\
4000&	0.8884884 \\
4500&	0.9217262 \\
\hline
\end{tabular}
\end{center}
\caption{Correlation coefficient between times of gene flow estimated using haplotype and genotype data vs the true time of gene flow.}
\label{table:cor}
\end{table}
\pagebreak

%% file: map-error.tex
\section{Correcting for uncertainties in the genetic map}
\label{sec:me}
In this section, we show how uncertainties in the genetic lead to a bias in the estimates of the time of gene flow. We then show how we could correct our estimates assuming a model of map uncertainty. Our model characterizes the precision of a map by a single scalar parameter $\alpha$. We estimate $\alpha$ for a given genetic map by comparing the distances between a pair of markers as estimated by the map to the number of crossovers that span those markers as observed in a pedigree. We propose a hierarchical model that relates $\alpha$ and the expected as well as observed number of crossovers and we infer an approximate posterior distribution of $\alpha$ by Gibbs sampling. Finally, we show using simulations that this procedure is effective in providing unbiased date estimates in the presence of map uncertainties and we apply this procedure to estimate the uncertainties of the Decode map and Oxford LD-based map by comparing these maps to crossover events observed in a Hutterite pedigree.

\subsection {Correction}
We have a genetic map $\mathcal{G}$ defined on $m$ markers. Each of the $m-1$ intervals is assigned a genetic distance $g_i, i\in\{1,\ldots,m-1\}$. These genetic distances provide a prior on the true underlying (unobserved) genetic distances $Z_i$. A reasonable prior on each $Z_i$ is then given by 
\begin{equation}
Z_i \sim \Gamma\left(\alpha g_i,\alpha \right)
\label{eqn:gamma}
\end{equation}
where $\alpha$ is a parameter that is specific to the map. This implies that the true genetic distance $Z_i$ has mean $g_i$ and variance $\frac{g_i}{\alpha}$. So large values of $\alpha$ correspond to a more precise map. The above prior over $Z_i$ has the important property that $Z_1+Z_2 \sim \Gamma \left(\alpha (g_1+g_2),\alpha\right)$ so that $\alpha$ is a property of the map and not of the specific markers used.

Given this prior on the true genetic distances, fitting an exponential curve to pairs of markers at a given observed genetic distance $g$, involves integrating over the exponential function evaluated at the true genetic distances given $g$ \emph{i.e.},
\begin{equation}
\EE \left[\exp\left(-t_{GF} Z\right)\vert g\right]=\exp\left(-\lambda g\right)
\label{eqn:eq1}
\end{equation}
where $\lambda$ is the rate of decay of $\overline{D}(g)$ as a function of the observed genetic distance $g$ and can be estimated from the data in a straightforward manner and $t_{GF}$ denotes the true time of the gene flow.
It also easy to see that $\lambda$ will be a downward biased estimate of $t_{GF}$ (applying Jensen's inequality). 

We can use Equation~\ref{eqn:gamma} to solve for $t_{GF}$ (see Appendix~\ref{sec:appb} for details) as 
\begin{equation}
t_{GF} = \alpha \left( \exp\left(\frac{\lambda}{\alpha}\right) - 1\right)
\label{eq:correction}
\end{equation}

Thus, we need to estimate $\alpha$ for our genetic map to obtain an estimate of $t_{GF}$. As a check, note that for a highly precise map, $\alpha\gg \lambda$, we have $t_{GF}\approx \lambda$. 

\subsection {Estimating $\alpha$}
\begin{figure}[h]
\begin{center}
\includegraphics[height=.6\textheight]{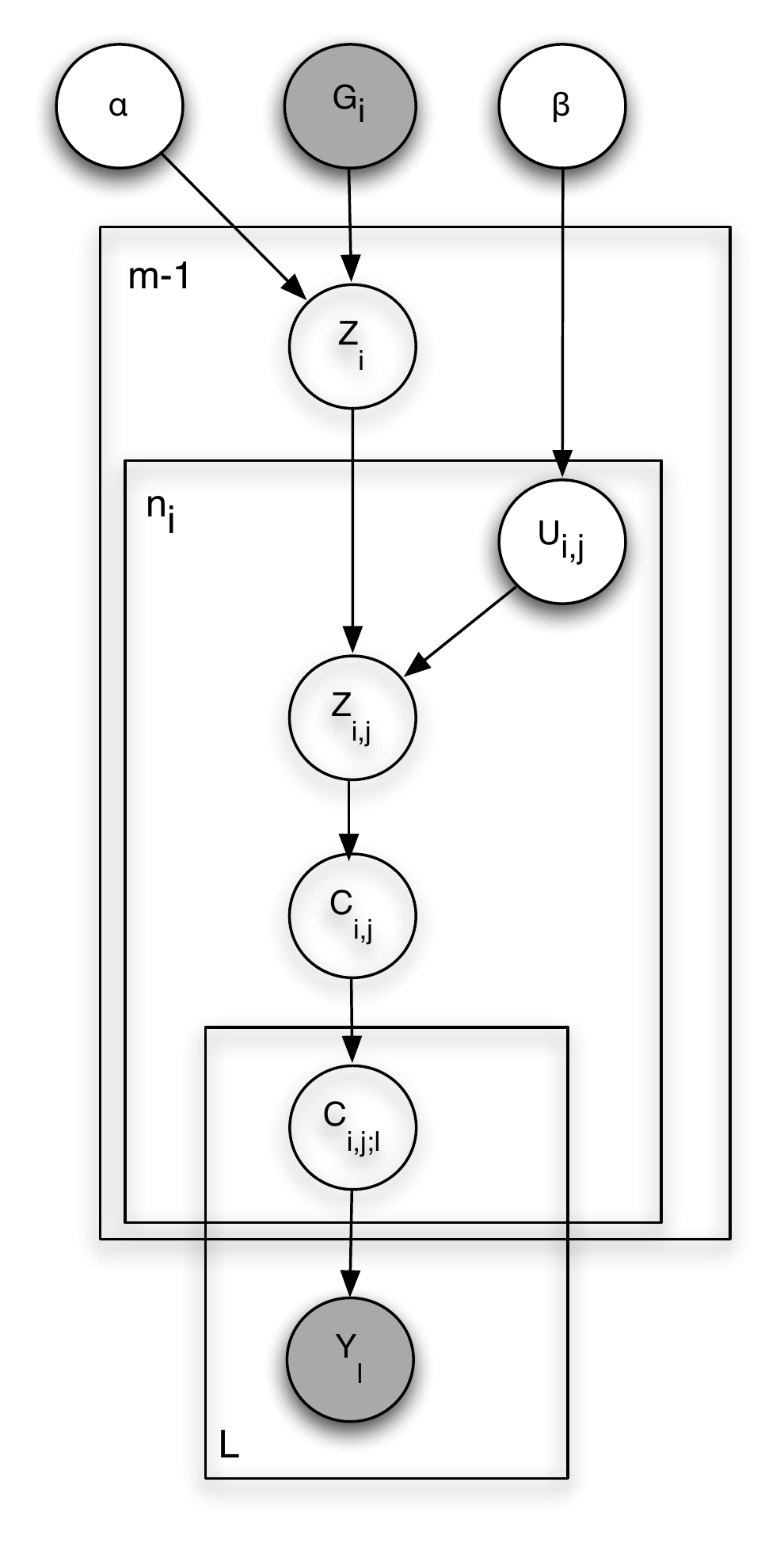}
\end{center}
\caption{A graphical model for map error estimation. Each circle denotes a random variable. Shaded circles indicate random variables that are observed. Plates 
indicate replicas of the random variables with the number of replicas denoted in the the top-left (\emph{e.g.}, there are $m-1$ copies of $Z_i$).$\alpha$ is the parameter that measures 
the precision of the map. $G_i, i \in [m-1]$ refers to the observed genetic distances across the $i^{th}$ interval in the genetic map. We impose an exponential prior on $\alpha$. $G_i$ and $\alpha$ parameterize the distribution over the true, but unobserved, genetic distance $Z_i$. $Z_i$ is gamma distributed with shape parameter $\alpha g_i$ and rate parameter $\alpha$. The genetic distance of interval $Z_i$ is partitioned amongst $[n_i]$ finer intervals to obtain genetic distances $Z_{i,j}$ using a Dirichlet distribution parameterized by $\beta$ and the physical distances of the finer intervals. Given $Z_{i,j}$, the number of crossovers $C_{i,j}$ within interval $(i,j)$ is given by a Poisson distribution with mean parameter $RZ_{i,j}$ where $R$ is the total number of meioses observed. These crossovers are then uniformly distributed amongst all the windows that overlap interval $(i,j)$. A crossover is observed within a window $l$, $Y_l=1$, only if one of the intervals spanned by this window is assigned a crossover.
}

\label{fig:maperror}
\end{figure}

Given a genetic map $\mathcal{G}$ defined on $m$ markers, each of the $m-1$ intervals is assigned a genetic distance $g_i,i \in [m-1]=\{1,\ldots,m-1\}$. Each interval $i$ may contain $n_i-1,\ge 0$ additional markers not present in $\mathcal{G}$ that partition interval $i$ into a finer grid of $n_i$ intervals -- each finer interval is indexed by the set $T=\{(i,j),i\in [m-1],j\in[n_i]\}$ (\emph{e.g.}, these additional markers could include markers that are found in the observed crossovers but not in the genetic map ). Each interval $(i,j)$ has a physical distance $p_{i,j}$.

We propose the following model for taking into account the effect of map uncertainty. 
\begin{eqnarray}
Z_{i}\vert\alpha,g_i &\sim& \Gamma(\alpha g_i,\alpha)\\
(Z_{i,1},\ldots,Z_{i,n_i})\vert U_i,Z_{i} &\sim& (U_{i,1}\ldots,U_{i,n_i})Z_i \\
U_i=(U_{i,1}\ldots,U_{i,n_i})\vert \beta &\sim& \Dir (\beta p_{i,1},\ldots, \beta p_{i,n_i})
\end{eqnarray}

The ``true'' genetic distance $Z_{i}$ is related to the observed genetic distance $g_i$ through the parameter $\alpha$ that is an estimate of map precision. The genetic distances of the finer intervals are obtained by partitioning the coarse intervals -- the variability of this partition is controlled by the parameter $\beta$ -- $\beta$ relates the physical distance to the genetic distance. When $\beta\rightarrow \infty$, the genetic distances of the finer grid are obtained by simply interpolating the coarse grid based on the physical distance. 

Given the true genetic distances, we can now describe the probability of observing crossovers. Our observed data consists of $R$ meioses that produce crossovers localized to $L$ windows $\{I_1,\ldots,I_{L}\}$. Each window $l \in [L]$ consists of a set of contiguous intervals $I_l$ and is known to contain a crossover event. Let $W_{i,j}$ denote the set of windows that overlap interval $(i,j)$.

A note on our notation: $C_{i,j;l}$ is the number of crossovers in interval $(i,j)$ that fall on window $l$. We can index the $C$ variables by sets and then we are referring to the total number of crossovers in the index set \emph{e.g.}, $C_{I_l;l}$ refers to all crossovers that fall on window $l$ within the set of intervals $I_l$. Omitting an index from a random variable implies summing over that index. Thus, $C_{i,j} = \sum_{l=1}^{L} C_{i,j;l}$ denotes the number of crossover events in interval $(i,j)$, $C_i = \sum_{j=1}^{n_i} C_{i,j}$ denotes the number of crossovers in the union of $(i,j), j \in [n_i]$. $\overrightarrow{.}$ indicates a vector of random variables $\emph{e.g.}$, $\overrightarrow{C}_{S}$ denotes the vector of counts indexed by the elements of set $S$.

%
If we assume that the probability of multiple crossovers in any of these intervals is small, we can use a simple probability model.
\begin{eqnarray}
C_{i,j}\vert Z_{i,j} &\sim& \Poi(R Z_{i,j})\\
\overrightarrow{C}_{i,j;l}\vert C_{i,j} &\propto&  \del{\sum_{\{l \in W_{i,j}\}}C_{i,j;l}\le C_{i,j},C_{i,j;l}\in \{0,1\}, \sum_{\{l \not \in W_{i,j}\}}C_{i,j;l} = 0 }\\
Y_l\vert C_{i,j;l} &=& \del{C_{I_l}=\sum_{(i,j)\in I_l} C_{i,j;l} =1}
\end{eqnarray}
Here $C_{i,j}$ denotes the counts of crossover events within interval $(i,j)$ over the $R$ meioses and is a Poisson distribution with rate parameter $RZ_{i,j}$. In our model, $C_{i,j;l}$ is either zero or one and all the crossovers in interval $(i,j)$ must fall on one of the $W_{i,j}$ windows that overlap $(i,j)$. Finally, one of the $C_{i,j;l}$ within a window $l$ must be one for a crossover to have been detected within this window ($Y_l = 1$).

We put an exponential prior on $\pi_{\alpha}\sim \exp(\frac{1}{\alpha_0})$ on $\alpha$. We set $\alpha_0 = 10$ in our inference.
While we can estimate $\beta$ jointly, we instead fix $\beta$ to $\infty$. 

To summarize, the observations in our model consist of the $m-1$ observed genetic distances $G_i,i \in [m-1]$ and $L$ observed crossovers from pedigree data $Y_l,l \in [L]$ (which often extend over multiple intervals in the underlying map) as well as the total number of meioses $R$ in the pedigree. The parameter of interest is $\alpha$, a measure of the precision of the map. We impose an exponential prior on $\alpha$. $G_i$ and $\alpha$ parameterize the distribution over the true, but unobserved, genetic distance $Z_i$. Given the number of meioses and $Z_i$, the number of crossovers that fall within interval $i$ (and is unobserved) is given by a Poisson distribution. These crossovers that fall within an interval $i$ are then distributed uniformly at random amongst all the observed windows that overlap interval $i$. Finally, a crossover is observed only if one of the intervals spanned by it is assigned a crossover. Our model can also account for the fact that the genetic map has been estimated using only a subset of markers from a finer set of markers (so that the markers defining the map and those defining the crossover boundaries may be different): the genetic distance of interval $Z_i$ is partitioned amongst the finer intervals $[n_i]$ to obtain genetic distances $Z_{i,j}$ using a Dirichlet distribution parameterized by $\beta$ and the physical distances of the finer intervals; given these $Z_{i,j}$, we can again compute the probability of observing a crossover across these finer intervals.

Thus, we are interested in estimating the posterior probability $\pi(\alpha \vert \overrightarrow{Y}, \overrightarrow{G}, \beta)$ where $\overrightarrow{Y} = (Y_1,\ldots,Y_{L})$, $\overrightarrow{G} = (G_1,\ldots,G_{m-1})$. $\pi(\alpha \vert \overrightarrow{Y}, \overrightarrow{G}, \beta) \propto \pi_{\alpha}(\alpha) \Pr(\overrightarrow{Y} \vert \alpha, \beta, \overrightarrow{G})$ where the likelihood is given by the probability model described above. To perform this inference, we set up a Gibbs sampler to estimate the posterior probability over the hidden variables $\pi(\alpha,\overrightarrow{Z}_{[m-1]},\overrightarrow{U}_{[m-1]},\overrightarrow{C}_{T}\vert \overrightarrow{Y}, \overrightarrow{G}, \beta)$.
\subsection{Inference}
We perform Gibbs sampling to estimate the approximate posterior probability over the hidden variables $(\alpha,\overrightarrow{Z}_{[m-1]},\overrightarrow{U}_{[m-1]},\overrightarrow{C}_{T})$. While a standard Gibbs sampler can be applied to this problem, mixing can be improved using the fact that we are interested in the estimates of $\alpha$ while the $Z_i$ are nuisance parameters. We thus attempt to sample $\alpha$ given the $C_{i,j}$, integrating out the $Z_i$. We still need the $Z_i$ in the model as it decouples the $C_{i,j}$. After sampling $\alpha$, we resample the $Z_i$ given the $\alpha$ and then resample $C_{i,j}$ given the resampled $Z_i$. 

Given the parameter estimates at iteration $t-1$, their estimates at time $t$ are given by
\begin{eqnarray*}
\Pr(\attime{\alpha}{t}\vert \attime{\overrightarrow{C}_{i}}{t-1}) &\propto& 
\prod_i \left(\frac{\Gamma\left( \attime{\alpha}{t}g_i + c_i \right)}{\Gamma \left( \attime{\alpha}{t} g_i\right)} \frac{{\attime{\alpha}{t}}^{\attime{\alpha}{t}g_i} }{ {\left( \alpha+R \right)}^{c_i+\attime{\alpha}{t} g_i} } \right) \exp\left(-\frac{\alpha}{\alpha_0}\right)\\
\attime{Z_{i}}{t}\vert \attime{\alpha}{t},\attime{C_{i}}{t-1}&\sim& \Gamma \left(\attime{\alpha}{t} g_i + \attime{C_{i}}{t-1}, \attime{\alpha}{t}+R\right)\\
\attime{U_{i}}{t} \vert \beta,\attime{\overrightarrow{C}_{i,[n_i]}}{t-1}&\sim& \Dir\left(\attime{\overrightarrow{C}_{i,[n_i]}}{t-1}+\beta \overrightarrow{p}_{i,[n_i]}\right)\\
\attime{Z_{i,j}}{t}\vert \attime{U_i}{t},\attime{Z_i}{t} &=& \attime{U_{i,j}}{t} \attime{Z_i}{t}
\end{eqnarray*}

In this sampler, $Z_{i,j}$ is a deterministic function of $Z_i$ and $C_i$, so we can collapse $Z_{i,j}$.

The first equation samples $\alpha$ given the current estimates of the counts $C_i$. This is not a standard distribution. We sample from this distribution using an ARMS sampler~\cite{arms}.

The genetic distances between the markers in the original map $\overrightarrow{Z}_i$ is a gamma distribution with parameters updated by $\attime{C_i}{t-1}$. The genetic distances between the markers in the finer grid $Z_{i,j}$ can now be obtained by sampling the $U_i$ which is a Dirichlet distribution with parameters updated by $\attime{C_i}{t-1}$.

We finally need to resample the counts $C_{i,j}$. For each window $l$, we can sample the total counts that fall within the window given the genetic distance spanned by the window (which in our simplified model is always $1$ for each window). We then assign each of these counts to one of the intervals within this window according to a multinomial distribution with probabilities proportional to their genetic distances. Finally $C_{i,j}$ is obtained by summing over the counts across all windows $W_{i,j}$ that overlap interval $(i,j)$.

\begin{eqnarray*}
\Pr(\attime{C_{I_l;l}}{t} \vert Y_{l}=1,\attime{Z_{I_l}}{t} )  &=&  \delta(C_{I_l;l}=1)\\
\attime{C_{i,j;l}}{t}\vert \attime{C_{I_l;l}}{t},\attime{Z_{I_l}}{t} &\sim& Mult\left(1, \attime{Z_{I_l}}{t}\right)\\
\attime{C_{i,j}}{t} \vert \attime{C_{i,j;l}}{t} &=& \sum_{l \in W_{i,j}} \attime{C_{i,j;l}} {t}\\
\attime{C_i}{t} \vert \attime{C_{i,j}}{t} &=& \sum_{j=1}^{n_i} \attime{C_{i,j}}{t}
\end{eqnarray*}

\subsection {Simulations}
To investigate the adequacy of our model of map errors, we performed coalescent simulations using a hotspot model of recombination. We estimated the time of gene flow using an erroneous map. We then estimated the uncertainty of the parameter $\alpha$ by comparing the erroneous map to the true genetic map. We used the estimated $\alpha$ to obtain a corrected date. This procedure allows us to evaluate if our model can capture the uncertainties in the genetic map.

We simulated $100$ independent $1$ Mb regions using MSHOT~\cite{mshot}. We chose parameters for the recombination model similar to the parameters described in ~\cite{hellenthal.pgen.2009}. We considered a model with $t_{NH}=10000,t_{YE}=5000,t_{GF}=2000$, constant effective population sizes of $10000$ and a bottleneck in the Neandertal lineage of duration $200$ generations and effective population size $100$. Given the true genetic map for each locus, the observed map is a noisy version generated as follows: given the genetic map length $l$ of each locus, the observed map has a genetic length $G$ distributed according to a Gamma distribution $\Gamma(al,a)$ where $a$ parameterizes the variance of the map \footnote{Note that $a$ is not the same as the parameter $\alpha$ that characterizes the variance of the true map given the observed map. $a$ parameterizes the variance in an observed map given the true map while $\alpha$ parameterizes the variance in the true map given an observed map}. Given $G$, the distances of the markers are obtained by interpolating from the physical positions. 

We obtained an uncorrected estimate of the date $\lambda$ using the observed genetic map. We then compared the true genetic map and the observed map to estimate $\alpha$ (restricting to markers at distances of at least $0.02$ cM ) and then obtained the corrected date $t_{GF}$ according to Equation~\ref{eq:correction}. Table~\ref{table:me} reports the results averaged over $10$ random datasets. We see that the corrected date $t_{GF}$ is quite accurate when the map is accurate at a scale of $1$ Mb ($a \ge 1000$) and becomes less accurate when $a\le 100$. The results are similar when we repeated the simulations with a demography in which there is a $20$ generation bottleneck of $N_e=100$ after the gene flow.

\begin{table}
\begin{center}
\begin{tabular}{lllll}
\hline
$a$&\multicolumn{2}{c}{No bottleneck since gene flow}&\multicolumn{2}{c}{Bottleneck}\\
&$\lambda$&$t_{GF}$&$\lambda$&$t_{GF}$\\
\hline
$\infty$&1597$\pm$180&1926$\pm$252&1660	$\pm$130&2005$\pm$194\\
$1000$&1653$\pm$198&2050$\pm$288&1715$\pm$127&2128$\pm$156\\
$100$&788$\pm$352&993$\pm$543&681$\pm$200&802$\pm$256\\
\hline
\end{tabular}
\end{center}
\caption{Estimates of time of gene flow as a function of the quality of the genetic map: Data was simulated under a hotspot model of recombination. The observed genetic map was obtained by perturbing the true genetic map at a 1 Mb scale and then interpolating based on the physical positions of the markers. Smaller values of $a$ indicate larger perturbation. $\lambda$ denotes the estimates obtained on the perturbed map. $t_{GF}$ denotes the estimates obtained after correcting for the errors in the observed map. Results are reported for two demographic models.}
\label{table:me}
\end{table}
\subsection {Results}
The previous results provide us confidence that the statistical correction for map uncertainty gives accurate estimates of the date provided the genetic map is reasonably accurate at a scale of $1$ Mb. In our analyses, we therefore chose to use the Decode map~\cite{kong.nature.2010} as well as the Oxford LD-based maps~\cite{myers.science.2005} which are known to be accurate at this scale. Another map that we considered using was a map obtained by using the physical positions to interpolate genetic distances estimated across entire chromosomes or sub-regions (e.g. the long arm, the centromere and the short arm). We chose not to use such a ``physical'' map because of its large variance at smaller size scales -- \emph{e.g.}, comparing this physical map to the Decode map suggests that the uncertainty in the genetic map is characterized by $a \approx 150$.

We estimated the uncertainty $\alpha$ of two maps -- the Decode map and the CEU Oxford LD map. In each case, we assigned genetic distances to the SNPs in the 1000 genomes CEU data. Our observed crossovers consisted of the crossovers observed in a family of Hutterites~\cite{coop.science.2008}. We ran our Gibbs sampler for $500$ iterations preceded by $250$ iterations of burn-in (even though the mixing happens much faster). We initialized $\alpha$ from the prior. Different random initializations do not affect our results (even though this is not a diagnostic for problems with the chain or bugs). Our estimates show that the precision of the CEU LD map and the Decode map  are quite similar with the Decode map being a little more accurate (see Table~\ref{table:alpha}).
\begin{table}[h]
\begin{center}
\begin{tabular}{ll}
\hline
Map&$\alpha$\\
\hline
Decode&1399.3$\pm$99.733\\
CEU&1221.89$\pm$78.79\\
\hline
\end{tabular}
\end{center}
\caption{Estimates of the precision of two Genetic maps}
\label{table:alpha}
\end{table}

%% file: uncertainty.tex
\section{Uncertainty in the date estimates}
\label{sec:unc}
We obtain estimates of the time of gene flow taking into account all sources of uncertainty. Denote the uncorrected date, the corrected date in generations and the corrected date in years by $\lambda$, $t_{GF}$ and $y_{GF}$ respectively. 

Our model can be described as follows:
\begin{eqnarray*}
t_{GF}&=&y_{GF} G \\
\lambda &=& \alpha \left(\log \left(\frac{t_{GF}}{\alpha}\right) + 1 \right)\\
\overline{D}(x) &=& a \exp \left(-\lambda x\right) + \epsilon  \\
\epsilon &\sim& N(0,\sigma^2)\\
\pi(\sigma^2) &\propto& \frac{1}{\sigma^2}
\end{eqnarray*}
where $G\sim Unif(25,33)$ denotes the number of years per generation, $\alpha$ is the uncertainty in the genetic map with prior given by the posterior  
estimated in Section~\ref{sec:me} and $a \sim Unif(0,1)$. Given this model, we can obtain the posterior probability distribution $\pi(\lambda \vert \overline{D}),\pi(t_{GF}\vert \overline{D}),\pi(y_{GF}\vert \overline{D})$ assuming a flat prior on each of the random variables $\lambda, t_{GF}, y_{GF}$ respectively.

We obtain these posterior distributions by Gibbs sampling. We ran the Gibbs sampler for $200$ burn-in iterations followed by $1000$ iterations where we sampled every $10$ iterations. We computed the posterior means and $95\%$ credible intervals on $\lambda, t_{GF}$ and $y_{GF}$. 

%% file: ascertain.tex
\section{Effect of ascertainment}
\label{sec:ascertain}
To test the robustness of our statistic, we performed coalescent-based simulations under the demographic models described in Section~\ref{sec:sims}. We explored two SNP ascertainments in addition to the ascertainment that we described in Section~\ref{sec:statistic} (which we refer to here as Ascertainment 0):
\begin{enumerate}
\item Ascertainment 1: SNPs for which Neandertal carries a derived allele, $E$ is polymorphic and $Y$ does not carry a derived allele.
\item Ascertainment 2: SNPs for which Neandertal carries a derived allele, $E$ is polymorphic and $Y$ does not carry a derived allele and SNPs for which Neandertal carries a derived allele, $E$ does not carry a derived allele and $Y$ is polymorphic.
\end{enumerate}
\subsection{Recent gene flow}
\begin{figure}
\begin{center}
\includegraphics[height=.4\textheight]{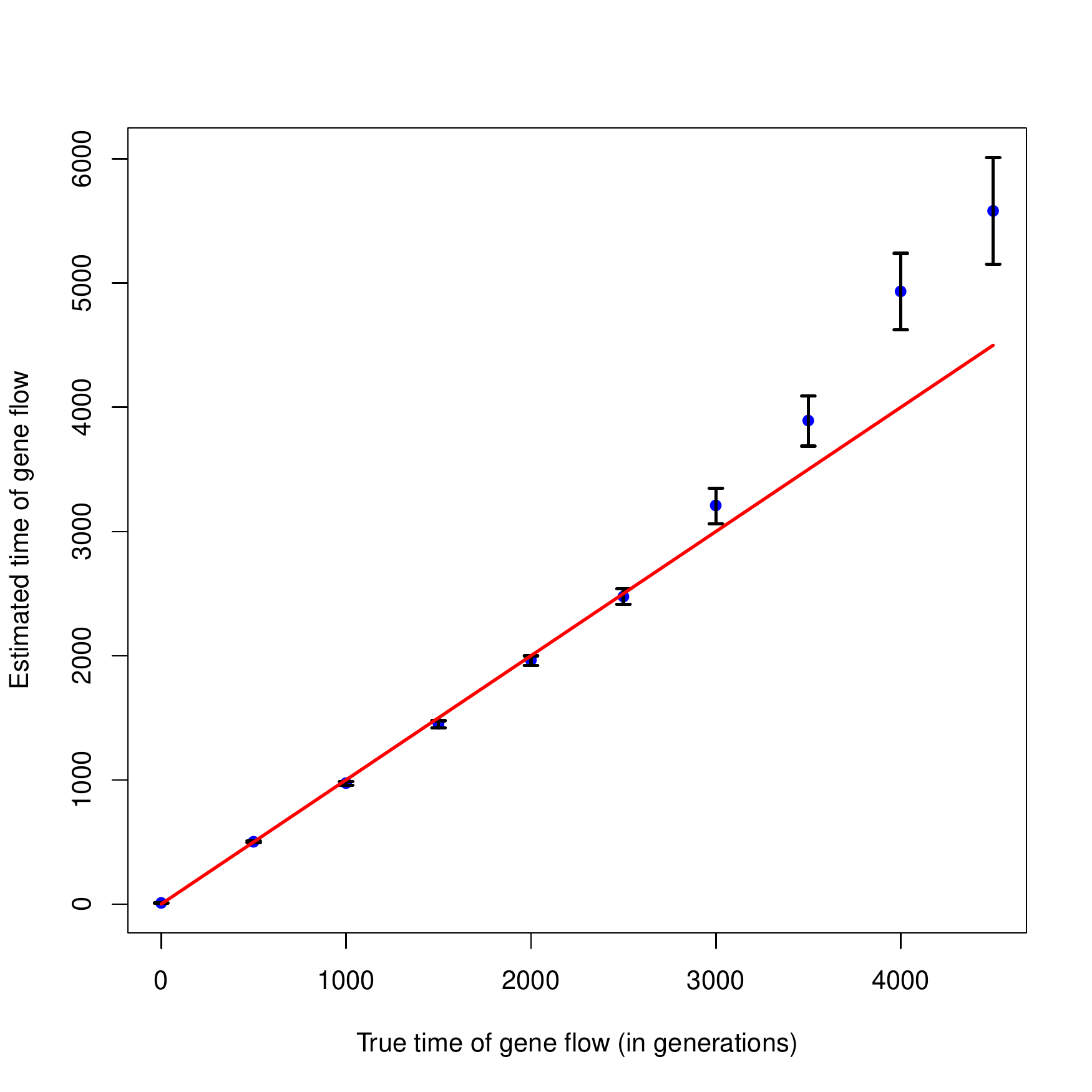}
\end{center}
\caption{Estimates of $t_{GF}$ as a function of true $t_{GF}$ for Demography RGF I: We plot the mean and $2\times$ standard error of the estimates of $t_{GF}$ from $100$ independent simulated datasets using ascertainment 1. The estimates track the true $t_{GF}$ though the variance increases for more ancient gene flow events.}
\label{fig:date-simple}
\end{figure}

\begin{figure}
\begin{center}
\includegraphics[height=.4\textheight]{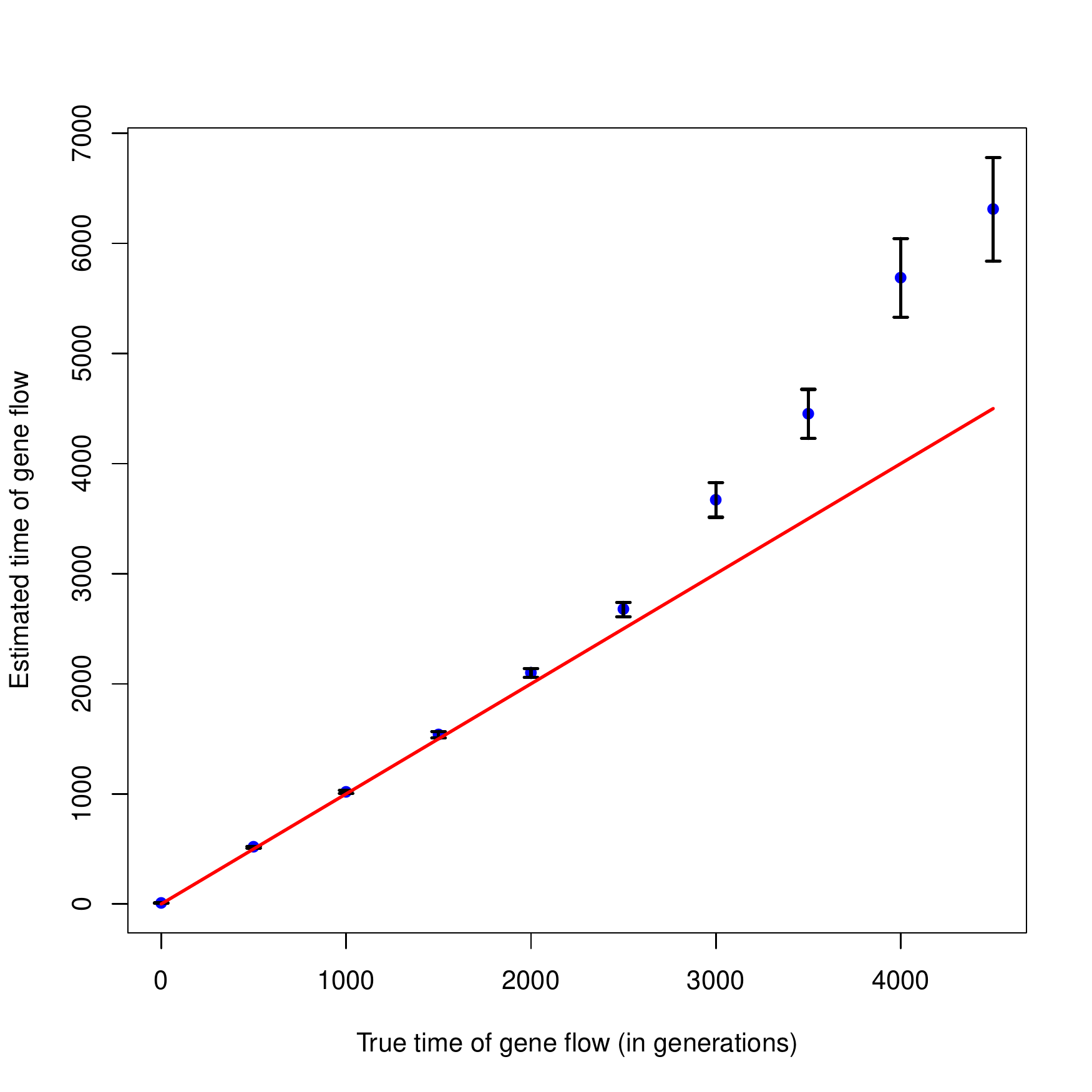}
\end{center}
\caption{Impact of the ascertainment scheme on the estimates of $t_{GF}$ as a function of true $t_{GF}$ for Demography RGF I: We plot the mean and $2\times$ standard error of the estimates of $t_{GF}$ from $100$ independent simulated datasets using ascertainment 2. }
\label{fig:date-complex}
\end{figure}

Under the simple demography I, Figures ~\ref{fig:date-simple} and ~\ref{fig:date-complex} show that, similar to ascertainment 0, the estimated $t_{GF}$ tracks the true $t_{GF}$ across the range of values of $t_{GF}$ for ascertainments 1 and 2. 

We assessed the effect of demographic changes since the gene flow on the estimates of the time of gene flow (demography RGF II of Section~\ref{sec:sims}). We see in Table~\ref{table:ascertain} that the bottleneck causes a downward bias in the estimated time using ascertainment $1$ while ascertainment $2$ is unbiased. 
For demography RGF III, Table~\ref{table:ascertain} shows that ascertainment $1$ again has a downward bias on the estimated date while ascertainment $2$ has a smaller upward bias.

\subsection{Ancient structure}
In the AS I model, ascertainments 1 and 2 both produce estimate close to the time of last gene exchange ($9000$ generations) as does ascertainment 0. In AS II, however, both ascertainments are less affected by the recent bottleneck in population $E$ and estimate older times that are closer to the true time of last gene exchange.

\subsection {No gene flow}
Both ascertainments 1 and 2 produce dates that are quite old for both models NGF I and NGF II -- the dates for NGF II are older than the estimates obtained using ascertainment 0. Ascertainment 2 produces estimates that are quite close to the time of last gene flow ($t_{NH}$).

Our simulation results show that in the case of recent gene flow, ascertainment 1 experiences a significant downward bias whereas ascertainment 2 is quite accurate. In the absence of gene flow or in the case of ancient structure, both ascertainments produce estimates that are quite old and they are more robust to population size changes in the target population relative to ascertainment 0.

\subsection {Hybrid Models}
For all the hybrid models, we see that all the ascertainments are quite accurate with ascertainment 1 being most accurate while ascertainments 0 and 2 have a small upward bias.

\subsection{ Effect of the mutation rate}
Mutation rate has an indirect effect on our estimates -- the mutation rate affects the proportion of ascertained SNPs that are likely to be introgressed. We varied the mutation rate to $1\times10^{-8}$ and $5\times10^{-8}$ in the RGF II model with no European bottleneck and again obtained consistent estimates (Table~\ref{table:ascertain}).

\subsection {Application to 1000 genomes data}
Due to the process of SNP calling that calls SNPs separately in each population, SNPs called in one of the populations may not have calls in another. This is particularly problematic for SNPs that are polymorphic in one population and monomorphic in the other -- precisely the SNPs that we would like to ascertain in the ascertainment schemes that we described above. To overcome this limitation, we used the following procedure to select our SNPs.
For each of the SNPs that are polymorphic in the target population, we estimated the allele frequencies in the ancestral population directly from the reads that mapped to the SNP.
We chose all SNPs whose derived allele frequency in the ancestral population is estimated to be less than $1\%$ (since we have $118$ YRI chromosomes, we can resolve frequencies of the order $\frac{1}{118} \approx 0.01$).

The ancestral allele, which was inferred using the Ensembl EPO alignment, was acquired from the 1000 Genomes Project FTP site.
To derive the allele frequencies, we downloaded the pilot-phase alignments from the same FTP.
We first adjusted each read alignment to avoid potential artifacts caused by short sequence insertions and deletions (INDELs),
and then estimated the allele frequency by maximizing the likelihood using an estimation-maximization (EM) algorithm.
More exactly, given we know the frequency $\phi^{(t)}$ at the $t$-th iteration, the estimate for the next round is:
$$
\phi^{(t+1)}=\frac{1}{2n}\sum_{i=1}^n\frac{\sum_{g=0}^2g\mathcal{L}_i(g)f(g;2,\phi^{(t)})}{\sum_{g=0}^2\mathcal{L}_i(g)f(g;2,\phi^{(t)})}
$$
where $n$ is the total number of samples, $f(g;2,\psi)=\binom{2}{g}\psi^g(1-\psi)^{2-g}$ is the frequency of genotype $g$ under the Hardy-Weinberg equilibrium,
and $\mathcal{L}_i(g)$ is the likelihood of $g$ for the $i$-th sample. The genotype likelihood $\mathcal{L}_i(g)$
was computed using the MAQ error model~\cite{li.gr.2008}.

The estimates of these different ascertainments are shown in Table~\ref{table:ascertain}. We observe that, as in the simulations, the estimates obtained using ascertainment 1 are lower than the dates obtained 
using ascertainment 0 while those using ascertainment 2 are closer.

Finally, we also considered the effect of the frequency threshold of $0.10$ used in Ascertainment 0. Using thresholds of $0.05$ and $0.20$, we obtain estimates of $\lambda=1201(1172,1233), 1188 (1164,1211)$ respectively using the Decode map. Thus, our estimates are not sensitive to the specific threshold chosen.

\begin{table}
\begin{center}
\begin{tabular}{llllll}
\hline
Demography&Ascertainment 0&Ascertainment 1&Ascertainment 2\\
\hline
RGF II& 1987$\pm$48& 1693$\pm$ 39& 1960$\pm$ 43\\
RGF III&1776$\pm$87&1642$\pm$98&2272$\pm$102\\
RGF IV &2023 $\pm$ 56 &1751$\pm$36 & 1995 $\pm$ 38 \\
RGF V & 2157$\pm$22 & 2094 $\pm$ 22 & 2105 $\pm$ 22\\
RGF VI & 2102$\pm$36 & 1814 $\pm$ 35 & 2029 $\pm$ 38\\
AS I& 10128$\pm$127 &8162$\pm$107&8861$\pm$110\\
AS II& 5070$\pm$397&6349$\pm$327&7570$\pm$433\\
NGF I &8847$\pm$126&7940$\pm$257&10206$\pm$280\\
NGF II& 5800$\pm$ 164&7204$\pm$   356&11702$\pm$  451\\
HM I&2174$\pm$40&2057$\pm$36&2228$\pm$38\\
HM II&2226$\pm$39&2049$\pm$30&2100$\pm$30\\
HM III&2137$\pm$34&2040$\pm$29&2124$\pm$30\\
HM IV&2153$\pm$36&2038$\pm$34&2187$\pm$35\\
\hline
Mutation rate & Ascertainment 0&Ascertainment 1&Ascertainment 2\\
$1\time 10^{-8}$&2141$\pm$41&1847$\pm$35&1969$\pm$36\\
$5\times10^{-8}$&2134$\pm$41&1833$\pm$29&1951$\pm$29\\
\hline
\end{tabular}
\end{center}
\caption{Estimates of time of gene flow for different demographies. For the demographies that involve recent gene flow (RGF II, RGF III, RGF IV and RGF V), the true time of gene flow is $2000$ generations.}
\label{table:ascertain}
\end{table}

\pagebreak

\section {Effect of the 1000 genomes SNP calling}
\begin{figure}
\begin{center}
\includegraphics[height=.4\textheight]{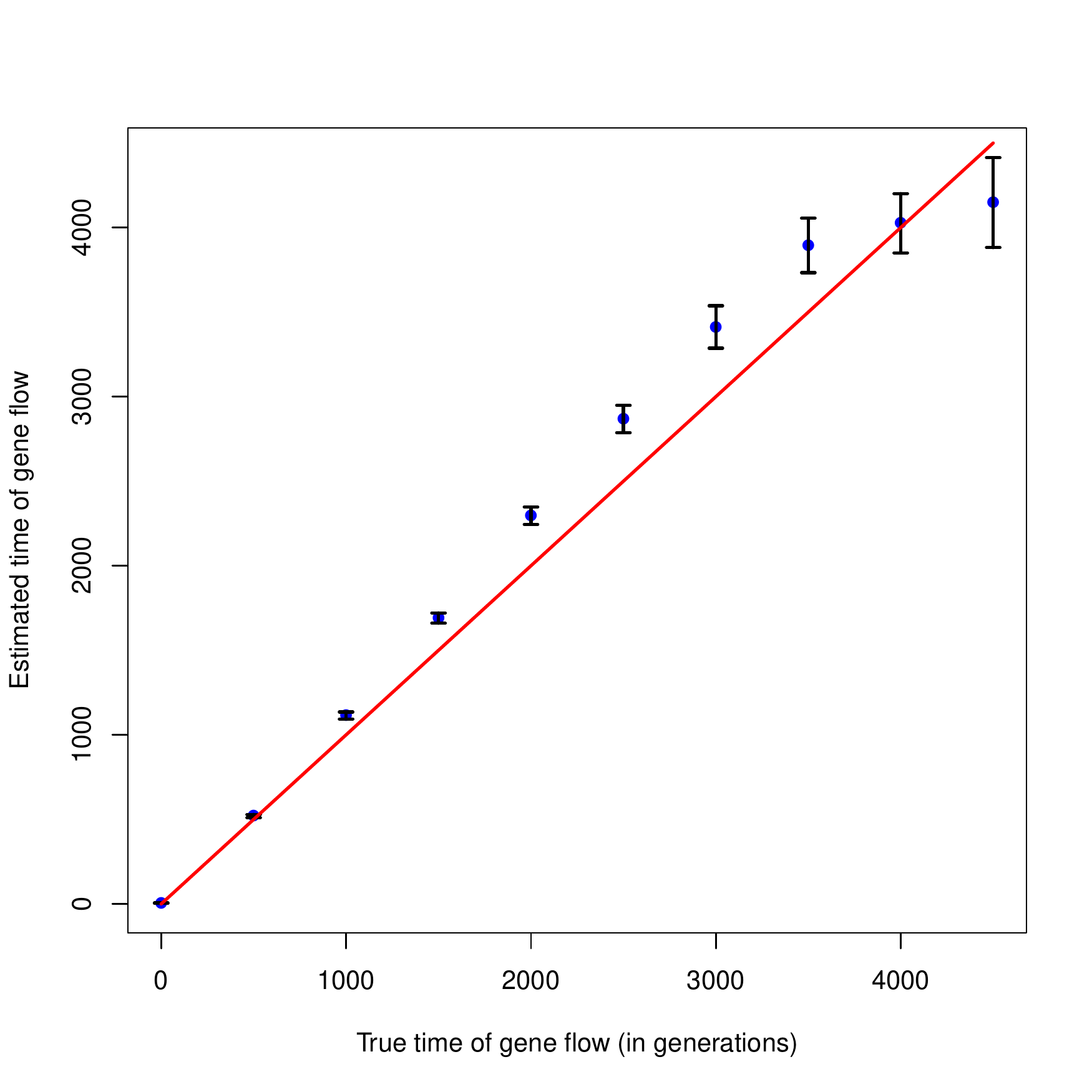}
\end{center}
\caption{Estimates of $t_{GF}$ as a function of true $t_{GF}$ for RGF I when the SNPs were filtered to mimic the 1000 genomes SNP calling process: We plot the mean and $2\times$ standard error of the estimates of $t_{GF}$ from $100$ independent simulated datasets using ascertainment 0. The estimates track the true $t_{GF}$ and are indistinguishable from estimates obtained on the unfiltered dataset as seen in Figure~\ref{fig:date-simple-0-1}.}
\label{fig:date-1kg}
\end{figure}

One of the concerns with the estimates obtained from SNPs called in 1000 genomes arises from the low power to detect low-frequency alleles. To assess the effect of missing low-frequency variants on our inference, we redid the simulations in the RGF I model where SNPs were filtered to mimic the 1000 genomes SNP calling. Each SNP was retained in the dataset as a function of the number of copies of the minor allele -- the acceptance probabilities are $0.25, 0.5, 0.75, 0.80, 0.9,0.95,0.96,0.97,0.98,0.99$ for minor allele counts of $1,2,3,4,5,6,7,8,\ge 9$ respectively. Figure~\ref{fig:date-1kg} shows that the estimates on this filtered dataset are indistinguishable from the unfiltered dataset showing that the low power to call rare alleles does not affect our inference.

\section {Effect of the 1000 genomes imputation}
A potential concern with interpreting our LD-based estimates applied to the SNPs called in 1000 genomes arises from the fact that genotype calling in the 1000 genomes project involves an imputation step which used LD in a reference panel to call genotypes~\cite{1kg}. It is unclear how this step affects our estimates. To understand the effect of imputation, we estimated the haplotype frequencies at pairs of SNPs directly from the 1000 genome reads aligned to the human reference hg18. We then used these haplotype frequencies to estimate LD (as opposed to the genotypic LD that we use in the rest of the paper)~\cite{lewontin}.

Similar to the estimate of allele frequencies from the sequencing data, the two-locus haplotype frequencies are
also estimated using an EM algorithm. Given $k$ loci,
let $\vec{h}=(h_1,\ldots,h_k)$ be a haplotype where $h_j$ equals 1 if
the allele at the $j$-th locus is derived, and equals 0 otherwise.
Let $\eta_{\vec{h}}$ be the frequency of haplotype $\vec{h}$ satisfying
$\sum_{\vec{h}}\eta_{\vec{h}}=1$, where
$$
\sum_{\vec{h}}=\sum_{h_1=0}^1\sum_{h_2=0}^1\cdots\sum_{h_k=0}^1
$$
Knowing the genotype likelihood at the $j$-th locus for the $i$-th individual $\mathcal{L}^{(j)}_i(g)$, we
can compute the haplotype frequencies iteratively with:
\begin{equation}\label{equ:hf}
\eta^{(t+1)}_{\vec{h}}=\frac{\eta_{\vec{h}}^{(t)}}{n}\sum_{i=1}^n\frac{\sum_{\vec{h}'}\eta_{\vec{h'}}^{(t)}\prod_{j=1}^k\mathcal{L}^{(j)}_i(h_j+h'_j)}
{\sum_{\vec{h}',\vec{h}''}\eta_{\vec{h'}}^{(t)}\eta_{\vec{h''}}^{(t)}\prod_{j}\mathcal{L}^{(j)}_i(h'_j+h''_j)}
\end{equation}

We restricted our analysis to SNPs chosen using ascertainment 0 and used the Decode map to determine our genetic distances. We fitted an exponential with an affine term to the decay curve to obtain an uncorrected date $\lambda = 1210$, consistent with $\lambda=(1179,1233)$ obtained using the genotypes called in 1000 genomes. Thus, the genotype imputation does not appear to be a major source of bias in our estimates. 

%% file: more-results.tex
\section{Results}
\begin{table}[h]
\begin{center}
\begin{tabular}{lllllll}
\hline
Map&\multicolumn{3}{c}{CEU}&\multicolumn{3}{c}{CHB+JPT}\\
&$\lambda$&$t_{GF}$&${{y}_{GF}}$&$\lambda$&$t_{GF}$&${{y}_{GF}}$\\
\hline
Decode &1201&1900&54540&--&--&--\\
&(1179,1233)&(1805,1993)&(47334,63146)&--&--&--\\
LD&1170&1961&56266&1269&--&--\\
&(1159,1183)&(1881,2043)&(49021,64926)&(1253,1287)&--&--\\
\hline
\end{tabular}
\end{center}
\caption{Estimated time of the gene flow from Neandertals into Europeans (CEU) and East Asians (CHB+JPT): $\lambda$ refers to the uncorrected time in generations obtained as described in Section~\ref{sec:statistic}. $t_{GF}$ refers to the time in generations obtained from $\lambda$ by integrating out the uncertainty in the genetic map as described in Section~\ref{sec:me}. ${{y}_{GF}}$ refers to the time in years obtained from $\lambda$ by integrating out the uncertainty in the genetic map and the uncertainty in the number of years per generation (We are reporting the posterior mean and $95\%$ Bayesian credible intervals for each of these parameters). Estimates of the time of gene flow were obtained for CEU using the Decode map and the CEU LD map. Estimates for CHB+JPT were obtained using the CHB+JPT LD map (We do not have a precise estimate of the uncertainty in this genetic map -- hence, we report only $\lambda$). }
\label{table1}
\end{table}

\begin{table}[h]
\begin{center}
\begin{tabular}{lllll}
\hline
&Map&\multicolumn{3}{c}{CEU}\\
&&$\lambda$&$t_{GF}$&${{y}_{GF}}$\\
\hline
Ascertainment 1 &Decode&962 &	1385&39760\\
&&(937,989)&(1328,1438)&(34593,45923)\\
&LD&1060&1694&48652\\
&&(1045,1074)&(1633,1755)&(42386,56065)\\
Ascertainment 2 &Decode&1105&1683&48311\\
&&(1080,1136)&(1590,1779)&(41796,56092)\\
&LD&1128 &1858 & 53332\\
&&(1089,1170)&(1764,1952)&(46134,61982)\\
\hline
\end{tabular}
\end{center}
\caption{Estimated time of the gene flow from Neandertals into Europeans (CEU) under different ascertainment schemes: $\lambda$ refers to the uncorrected time in generations obtained as described in Section~\ref{sec:statistic}. Ascertainment 1 is shown to have a downward bias in the presence of bottlenecks since the gene flow -- this may reflect the lower estimates obtained here. The estimates using Ascertainment 2 closely match the estimates shown in Table~\ref{table1}.}
\label{table2}
\end{table}

%
%
%
\begin{table}[h]
\begin{center}
\begin{tabular}{llllll}
\hline
Distance to exon&$\lambda$&$t_{GF}$&${{y}_{GF}}$\\
\hline
0-2475&1301&2149&61683\\
&(1256,1363)&(1991,2347)&(52737,72737)\\
2475-11028&1223&1967&56432\\
&(1176,1261)&(1874,2075)&(48708,65799)\\
11028-33707&1179&1847&53019\\
&(1131,1220)&(1717,1970)&(45679,61846)\\
33707-105107&1145&1773&50891\\
&(1098,1200)&(1640,1922)&(43330,59962)\\
105107-&1301&2151&61747\\
&(1253,1358)&(1982,2345)&(52442,73518)\\
\hline
\end{tabular}
\end{center}
\caption{Estimate of the time of gene flow stratified by distance to nearest exon (each bin contain $20\%$ of the $1000$ genome SNPs): These estimates were obtained on CEU using the Decode map. The results indicate that our estimates are not particularly sensitive to the strength of directional selection, which has recently been shown to be a widespread force in the genome~\cite{mcvicker.pgen.2009,cai.pgen.2009}. }
\label{table3}
\end{table}

\begin{figure}
\begin{minipage}{0.5\textwidth}
\centering
\includegraphics[width=\textwidth]{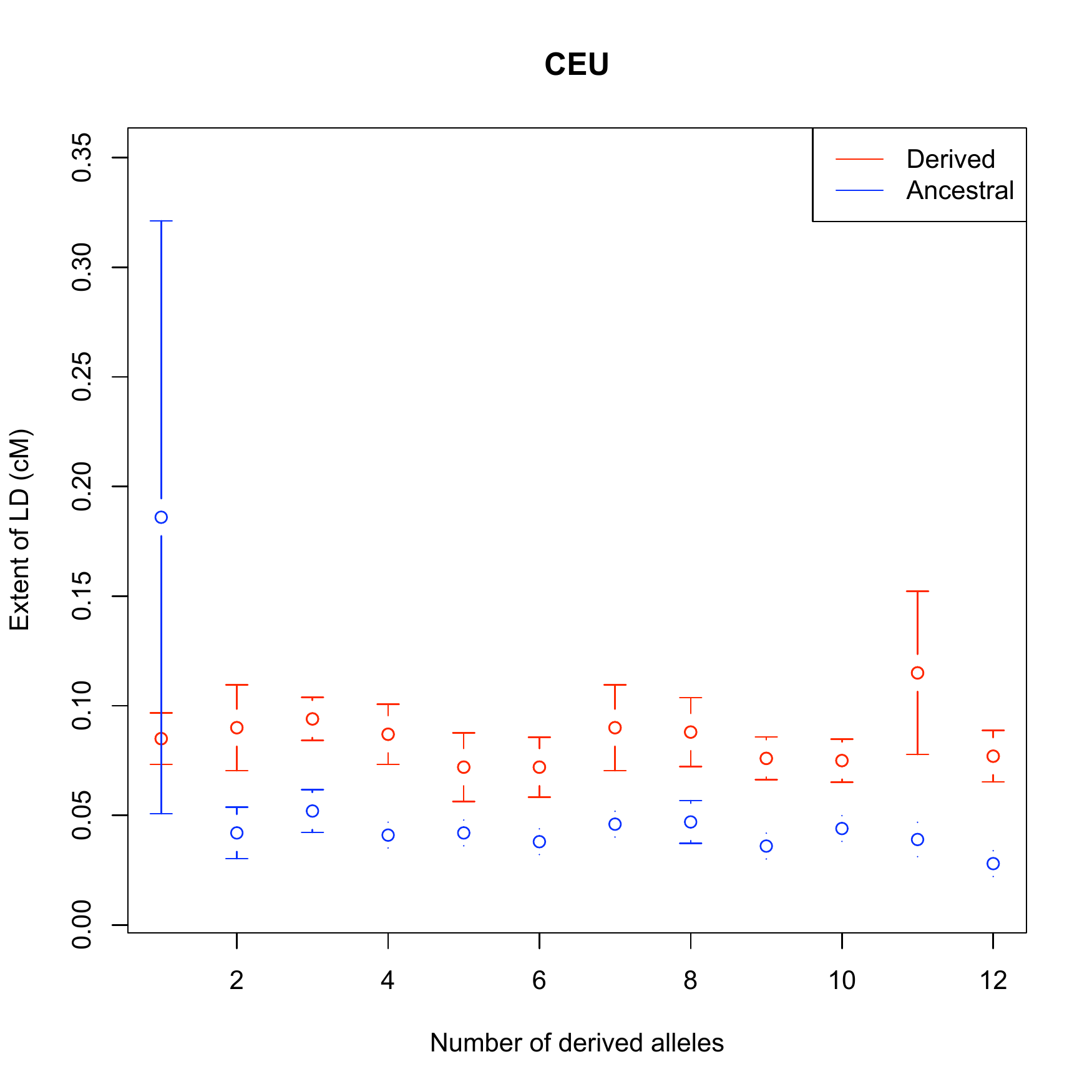}
\end{minipage}
\begin{minipage}{0.5\textwidth}
\centering
\includegraphics[width=\textwidth]{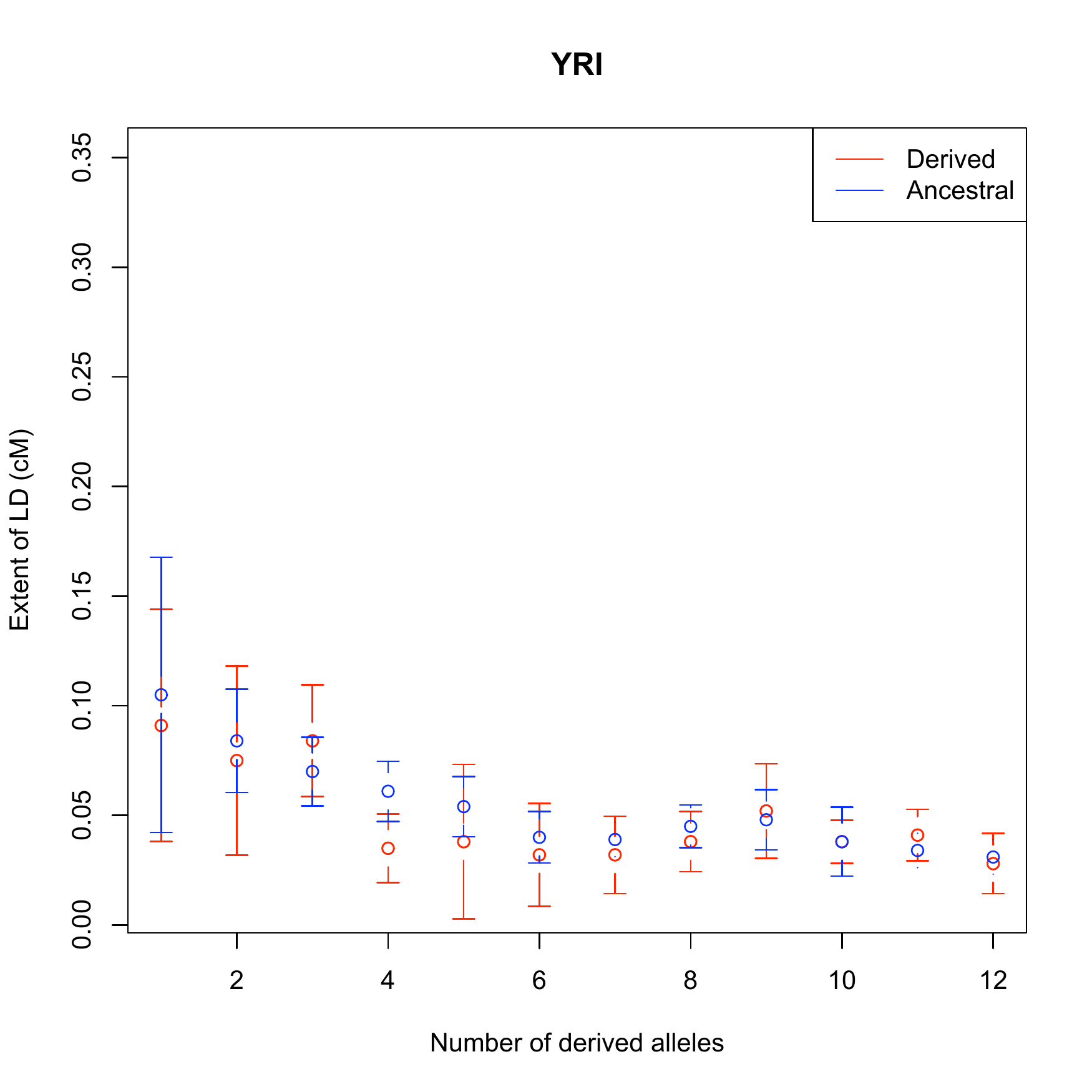}
\end{minipage}
\caption{ Comparison of the LD decay conditioned on  Neandertal derived alleles and Neandertal ancestral alleles stratified by the derived allele frequency in CEU (left) and YRI (right): In each panel, we compared the decay of LD for pairs of SNPs ascertained in two ways. One set of SNPs were chosen so that Neandertal carried the derived allele and where the number of derived alleles observed in the 1000 genomes CEU individuals is a parameter $x$. The second set of SNPs were chosen so that Neandertal carried only ancestral alleles and where the number of derived alleles observed in 1000 genomes CEU is $x$. We varied $x$ from $1$ to $12$ (corresponding to a derived allele frequency of at most $10\%$). For each value of $x$, we estimated the extent of the LD \emph{i.e.}, the scale parameter of the fitted exponential curve. Standard errors were estimated using a weighted block jackknife. Errorbars denote $1.96 \times $ the standard errors.  The extent of LD decay shows a different pattern in CEU vs YRI. In YRI, the extent of LD is similar across the two ascertainments to the limits of resolution although the point estimates indicate that the LD tends to be greater at sites where Neandertal carries the ancestral allele ($8$ out of $12$). In CEU, on the other hand, the extent of LD is significantly larger at sites where Neandertal carries the derived allele (the only exception consists of singleton sites). Thus, the scale of LD at these sites must be conveying information about the date of gene flow. } 
\label{fig:compare-ld}
\end{figure}

%% file: theory.tex
\section{Exponential decay of the statistic}
\label{sec:theory}

We are interested in how the linkage disequilibrium varies as a function of genetic distance $x$. We consider two SNPs that are polymorphic at time $0$ in the past. The evolution of the alleles at the two SNPs can be described by the two-locus Wright-Fisher diffusion in a space parameterized by $X_t=(p,q,D)_t$ \emph{i.e.}, the allele frequencies at each SNP at time $t$ and measure of LD $D$ at the two SNPs~\cite{ohta.genetics.1969}. At time $t$, the average LD is denoted $\EE D_{t}(x)$ (we assume that the population is not at equilibrium so that $\EE D \ne 0$). 

We are interested in the average linkage-disequilibrium at a time $t$ given the state of the system at time $0$ : $u(t,x) = \EM {D_{t}\vert X_0=x} $.  

We also denote the effective population size at time $t$ by $N(t) = \nu(t) N_0$ and the probability of recombination between the two loci by $r$.

The evolution of $u(t,x)$ is given by
\begin{equation}
\frac{\partial u}{\partial t} = \mathcal{L} u
\end{equation}
where $\mathcal{L}$ is the generator for this diffusion with initial condition 
\begin{equation*}
u(0,x) = D_0
\end{equation*}
and boundary conditions
\begin{eqnarray*}
u(t,(0,q,d)) = u(t,(p,0,d) &=& 0\\
\frac{\partial u}{\partial d}(p,q,d_{max}(p,q)) = \frac{\partial y}{\partial d} (p,q,d_{min}(p,q))&=& 0\\
\end{eqnarray*}
\begin{equation}
\frac{\partial u } {\partial t } =  \mathcal{L}u = - \left[ r + \frac{1}{2\nu(t)N_0} \right] u 
\label{eq:pde}
\end{equation}

The solution to Equation~\ref{eq:pde}  is given by 
\begin{eqnarray*}
u(t,x) &=& D_{0} \ef{-\integ{t}} \ef{ -rt  } 
\end{eqnarray*}
So we have 
\begin{eqnarray}
\EE D_t &=& \EE D_{0} \ef{-\integ{t}} \ef{ -rt  } 
\label{eq:main}
\end{eqnarray}

If we choose SNPs that that arose in the $N$ lineage and introgressed into $E$ $t_{GF}$ generations ago (\emph{i.e.}, these are SNPs that were monomoprhic in $E$ before the gene flow), Equation~\ref{eq:main} says that the average $D$ observed between all such pairs of SNPs at a given genetic distance $r$ depends on three factors -- the average LD at time $0$ ($\EE D_{0}$), the factor $\ef{-\integ{t}}$ that accounts for changes in population sizes since gene flow and the factor $\ef{-rt}$ that accounts for the decay in LD. Terms $1$ and $3$ depend on $r$ while term $2$ does not. Further, since we ascertain SNPs that arose in the $N$ lineage and introgressed into $E$, $\EE D_{0}$ will depend on the average value of $D$ in the introgressing Neandertals scaled by their admixing proportion. While $\EE D_{0}$ still depends on the genetic distance $r$, for highly-bottlenecked populations such as the Neandertals ,in which the probability of coalescence has been estimated to be at least $0.65$~\cite{reich.science.2010}, this term could be assumed to be a constant in $r$. We can then approximate the relation between the average $D$ and the genetic distance $r$ by the exponential term $\ef{-rt_{GF}}$ where the intercept of the exponential (its value at $r=0$) depends on the population history. 
Thus, rate of decay of the expectation of $D$ as a function of $r$ would correspond in this case to $t_{GF}$ and could provide a robust method to date gene flow.

Equation~\ref{eq:main} implies that changes in the effective population size since the gene flow will not change the relation between $\EE D_t$ and $r$. Since we have chosen SNPs that are monomorphic in $E$ before the time of gene flow, demographic history in $E$ before gene flow also does not affect $\EE D_t$.  However, this result has limitations when applied to polymorphism data. First, this result requires precisely ascertaining SNPs that arose in $N$ and introgressed. Imperfections in the ascertainment can make the procedure sentive to demography. Further, the expectation needs to be computed over all pairs of SNPs that were polymorphic at time $0$ even if these SNPs may have fixed or gone extinct since. Such SNPs would be hard to ascertain using present-day genomes. Second, if the drift since gene flow is high or the level of gene flow is low, the intercept of the exponential curve decreases making it harder to estimate its rate of decay from data

%% file: appb.tex
\section{Proof of Equation 3 in Section S3}
\label{sec:appb}
Equation ~\ref{eqn:eq1} is given by
\begin{equation}
\EE \left[\exp\left(-t_{GF} Z\right)\vert g\right]=\exp\left(-\lambda g\right)
\end{equation}
where 
\begin{equation}
Z \sim \Gamma\left(\alpha g,\alpha \right)
\end{equation}

We can explicitly compute the  LHS of ~\ref{eqn:eq1}
\begin{eqnarray}
\EE \left[\exp\left(-t_{GF} Z\right)\vert g\right] &=& \frac{\alpha^{\alpha g} }{\Gamma(\alpha g)}\int dz z^{\alpha g-1} \exp{\left(-\alpha z\right)}\exp\left(-t_{GF} z\right) \nonumber\\ 
&=& \frac{\alpha^{\alpha g} }{\Gamma(\alpha g)} \frac{\Gamma(\alpha g)}{{(t_{GF}+\alpha)}^{\alpha g} } \nonumber\\
&=& \exp \left( - \alpha \log \left (\left(\frac{t_{GF}}{\alpha}\right) + 1\right) g\right)
\label{eqn:eq2}
\end{eqnarray}
Equating the coefficients of $g$ in the RHS of Equation~\ref{eqn:eq1} and ~\ref{eqn:eq2} gives us Equation~\ref{eq:correction}.